# Engineering Collective Microbial Dynamics for Sustainable Thermal Management

Nilanjan Mondal[a†], Soumitree Mishra[a†] & Anupam Sengupta[a,b] *

[a]Physics of Living Matter, Department of Physics and Materials Science, University of Luxembourg

[b]Institute for Advanced Studies, University of Luxembourg

[†] These authors contributed equally to this work

*To whom correspondence should be addressed: anupam.sengupta@uni.lu

**Funding:** Funded by the European Union (ERC Consolidator Grant, MicroPAS, 101231075). Views and opinions expressed are however those of the authors only and do not necessarily reflect those of the European Union or the European Research Council Executive Agency. Neither the European Union nor the granting authority can be held responsible for them. This work received further support of the FNR Luxembourg's ATTRACT Investigator Grant (Grant no. A17/MS/11572821/ MBRACE to AS), the PRIDE Doctoral Training Unit ACTIVE (Active Phenomena Across Scales in Biological Systems, PRIDE19/14063202/ACTIVE), as well as funding from the University of Luxembourg.

**Author Contributions:** Conceptualization, planning, administration, and supervision: A.S. Methodology: N.M., S.M., and A.S. Investigation, data acquisition, mathematical analysis, and graphical representations: N.M. S. M., with inputs from A.S. Writing, review & editing: N.M., S.M., and A.S.

**Competing interests:** The authors declare that the research was conducted in the absence of any commercial or financial relationships that could be construed as a potential conflict of interest.

**Data and materials availability:** All relevant data supporting the findings are included in the manuscript. Additional datasets and analysis code (MATLAB & Python scripts for image analysis) are available from the corresponding author upon reasonable request.

## Abstract

The rapid growth of energy-intensive technologies, including artificial intelligence, large-scale computing infrastructure, and thermal management systems for buildings, industries, and public spaces, has intensified global energy demands amid accelerating climate change. This raises a critical question: can existing energy systems and environmental safeguards support future energy requirements without further ecological degradation? Addressing this challenge requires not only scalable computational hardware but also low-carbon energy sources and innovative thermal management strategies. This review revisits the underexplored phenomenon of bioconvection—the self-organized fluid motion generated by the collective dynamics of motile microorganisms—as a potential bio-inspired solution for sustainable heat transfer. Drawing on evidence from natural ecosystems and laboratory studies, we synthesize current knowledge of microorganism-induced hydrodynamics, pattern formation, and thermofluidic transport to evaluate the feasibility of harnessing bioconvection for sustainable thermal management. We further support this assessment with quantitative analyses to predict the thermal performance of such systems, and identify the relevant non-dimensional numbers. Motile microorganisms generate spontaneous convective plumes that significantly enhance heat and mass transfer without external mechanical forcing. By exploiting these self-organized micro- and mesoscale flow instabilities, bioconvection offers a promising route to hybrid bio-engineered cooling systems that reduce pumping energy, disrupt thermal boundary layers, and improve overall heat transfer efficiency. Realizing practical bioconvection-based cooling technologies will require advances in microbial stability, material compatibility, controllability of collective dynamics, scalability, and integration with existing thermal management infrastructure. This review summarizes the current state of bioconvection-mediated thermal transport, specifically in relation to the microbial traits, relevant engineering challenges, and outlines future research directions spanning heat transfer, microbiology, nonlinear fluid mechanics, within the broader context of sustainability. By critically assessing current understanding and emerging opportunities, we position bioconvection as a promising strategy for environmentally responsible thermal management in an era of rapidly increasing energy demand.

## 1. Introduction

As the world becomes increasingly reliant on energy-intensive technologies, the demand for efficient thermal management has grown dramatically across domestic, industrial, and digital infrastructures. Future projections indicate a continued and substantial increase in cooling and heat dissipation requirements driven by expanding computational capabilities and global electrification [1,2]. The widespread integration of artificial intelligence (AI), high-performance computing (HPC), quantum computing facilities, and high-density data centres is expected to increase annual cooling energy consumption to hundreds of terawatt-hours [1,3–6]. At the same time, accelerating climate change and global warming necessitate the development of sustainable cooling strategies that can be integrated across residential, commercial, and industrial sectors [7,8]. Efficient thermal management is equally critical for renewable energy technologies. For example, elevated operating temperatures substantially reduce the efficiency of solar cells [9], creating an additional demand for energy-intensive cooling solutions [10]. Consequently, considerable efforts have focused on developing passive cooling systems, advanced heat exchangers, and bio-inspired or microorganism-mediated heat transfer concepts. However, translating these laboratory-scale innovations into deployable technologies remains a major challenge [11].

Among emerging approaches, flow-mediated heat transfer using synthetic or living microswimmers has attracted growing interest owing to its ability to enhance mixing and redistribute heat without conventional mechanical pumping [12–14]. Similar challenges arise in modern bioreactors—including stirred-tank fermenters, photobioreactors, bubble columns, airlift reactors, and membrane systems—where dense, metabolically active cultures must be mixed efficiently while minimizing cellular damage [15,16]. In solar cooling applications, motile phytoplankton, whose collective swimming behavior follows diel vertical migration and responds to environmental cues, offer a unique opportunity to achieve self-regulated cooling synchronized with solar irradiance cycles. Harnessing such naturally occurring, self-organized biological flows, however, requires a deeper understanding of microbial dynamics and their coupling with heat transfer. This review highlights the potential of these collective microbial phenomena as a foundation for next-generation bio-hybrid and bio-inspired thermofluidic technologies applicable across diverse engineering sectors.

Recent reports highlight that almost 40 % of a data centre's electricity is used for cooling [17] and according to International Energy Agency (IEA) and S&P Global, the net electricity usage is anticipated to be more than double by 2030 driven by AI workloads [18,19]. In Europe, electricity supplied to data centres lies around 18.7 GW at the end of 2024 with a projection of 36 GW by 2030 [20,21]. In conventional cooling technologies such as immersion cooling, direct liquid cooling, or aggressive free cooling used today heavily rely on externally imposed gradients and mechanical actuation [18,22]. On the industrial side, markets for energy-efficient cooling systems are growing at 7–10% per year, reflecting simultaneously rising thermal loads and the cost pressure from their electricity use [23,24]. Enhancing heat transfer typically requires higher flow velocities, stronger vortical motions, larger temperature gradients, or increased interfacial area, all of which incur energetic costs through viscous dissipation and irreversible losses in systems from microscale electronics or data centre cooling to climate control in built environments [19,25,26]. Figure 1 highlights how the energy demand across diverse sectors will continue increasing over the coming years. The drivers of commercial growth are unimaginably high in future, which necessitates novel sustainable solutions that can save substantial energy in future.

Among the existing cooling technologies specially for data centers or high-performance computing, the first one we would like to mention is single- or two-phase immersion cooling where the servers or systems are fully submerged in a dielectric liquid with very high capacity and thermal conductivity [27–30]. The limitations in such systems mainly are the cost behind the liquid, dedicated pumping technologies/energies, custom tanks, sealed modules, fluid handling, redesign for racks, operation/equipment complexities, and service workflows [26,28,29,31]. The second conventional method is direct liquid cooling, and is limited largely by retrofitting issues, cost for liquids and pumping, and often close contact with electronics [32,33]. The third one that is the advanced free cooling/economisers/heat reuse using outside air, evaporative systems, lakes/groundwater, or district cooling [34,35]. Though they seem to be more sustainable at the first glance, this last technique is limited by factors such as strong climate/location dependence, risk of air quality/contamination/corrosion, water use and contamination with environments. Majority of the above-mentioned cooling methods except the last one involves forced convection or pumping and phase-change systems that rely on centralized energy input to maintain fluid motion or cyclic transitions between thermodynamic states [36,37]. Even the passive strategies ultimately depend on externally imposed temperature gradients and are bounded by diffusive transport limits. At reduced length scales, the energetic inefficiency becomes more severe: viscous stresses scale inversely with system size [38,39], while heat flux remains constrained by molecular transport [39,40]. These scaling limitations expose a fundamental shortcoming of conventional approaches: transport is driven at the macroscopic level, while dissipation is distributed throughout the system. Overcoming this mismatch requires rethinking how and where energy is injected into the transport process.

Active matter provides a route to fundamentally alter transport by shifting energy injection landscape from global to local scales [41]. In these systems, individual units convert internal or environmental energy directly into motion, giving rise to persistent nonequilibrium stresses and considerable collective flows [42,43]. One of the focused collective dynamics in microbial species observed in planktonic active matter systems is when a dense suspension of motile microalgae spontaneously generates convective mixing. When a sufficiently large population of motile microorganisms undergoes directed upward migration against gravity, preferential cell accumulation near the upper fluid layers establishes a gravitationally unstable density gradient. The emerging hydrodynamic instability generates descending plumes and complex spatiotemporal flow structures known as bioconvection, a self-organized collective transport phenomenon widely observed in both natural aquatic environments to controlled laboratory systems [43–47]. Theoretical models and experiments have shown that suspensions of motile gravitactic species can exhibit spontaneous circulation, enhanced mixing, and dynamic migratory strategies without any external energy input [46–49]. Such microbial flows and emergent hydrodynamic mixing can surely be harnessed for thermofluidic transport and management as a potential sustainable alternative for thermal management in future electronic components, data centres, and even industrial components. In case of bioreactors, temperature control represents a significant and frequently underappreciated fraction of bioreactor operating costs [50], particularly for enclosed bioreactors exposed to solar irradiation. Culture temperature control in PBRs is achieved either by passive-evaporative cooling such as spraying of freshwater on the surface [51,52], or by heat exchangers often limiting their large scale uses [52,53]. In larger bioreactors, the reduced surface-to-volume ratio makes it harder to remove heat, and thermal stratification is a persistent challenge [53]. Metabolic heat generation compounds solar heating: exothermic biochemical reactions produce a distributed volumetric heat source that creates internal temperature gradients even in the absence of external irradiation [34]. As a result of high light intensities, the temperature in photobioreactors can reach temperatures up to 50℃; control of temperature is therefore essential to avoid losses in biomass productivity but should be limited to a minimum to avoid high energy requirements for cooling [54,55]. Enhancement in environmental transport emerging from collective dynamics of motile microbes suggesting a qualitatively different mechanism from classical convection. Despite these advances, most studies have focused on synthetic swimmers or simplified biological models, and the implications for coupled heat and mass transport remain only partially explored.

By directly transducing light into metabolic energy and motility, photosynthetic microorganisms act as continuously powered, self-sustaining active agents. Their motility traits may largely vary from shorter to longer time span such as multiple generations depending on the flagellar beating, morphology, or intracellular configuration [47,56–58]. They adapt and develop strategies in order to survive with changing environmental conditions [59], flow fields [47,56], and light conditions [60] which possess countless non-trivial attributes and a promise at the same time towards developing future applications involving active matter [61]. The interaction between the microorganisms and the flow fields hide countless non-intuitive attributes directly influencing the shear fields along with the density or viscosity distributions due to their size, shape, livelihood, and distribution in their habitats [56,61,62]. While bioconvection has been long studied as a pattern-forming instability, its implications for heat and mass transport, particularly from a scaling and energy-efficiency perspective, have received comparatively little attention despite the observed variabilities in terms of flow interactions, time and length scales. This review is an endeavour to realise the unnoticed potentials of microbe collective mechanism in flow control and heat transfer mechanisms based on the previous research at the single cell to collective level behaviour. The goal is to bring a notice towards sustainable thermal management leveraging biological micro-swimmers and their collective traits to support fast transformation towards sustainable energy landscape in a changing world. Additionally, this review takes an attempt to realise the microbial active dynamics in terms of engineering perspective to lay a foundation that can guide the future experiments and numerical studies. It draws our attention towards the bioconvection as a physically grounded transport mechanism where the interrelations among the activity-induced flows, emergent time and length scales, and transport mechanism are yet to be revealed from multiple length scales. By systematically going through the previous experimental observations, proposed theoretical perspectives, and established scaling laws, we aim to bridge the gaps in this field and identify the regimes in which microbial activity can outperform passive transport with emergence of novel hybrid biobased coolants and to clarify the physical limits of such enhancement. Furthermore, this review positions photosynthetic microbes not merely as biological curiosities, but as model active systems that challenge conventional notions of how energy, flow, and transport are coupled in non-equilibrium fluids.

We first focus on the conventional and recent practices in flow-mediated heat transport where we discuss the passive and active non-living to living particles that have been introduced into the fields in Section 2. Moving forward, Section 3 shortly begins with bioconvection and how different factors along with the taxis mechanisms may influence the onset of the phenomena. We provide more attention towards bioconvection as a potential strategy towards thermal management due to its coherent and self-organized structures in various instances. With the promise of active microbial collective flows towards future thermofluidics including bioconvection, we revisit the theoretical frameworks applicable to predict and model the bioconvective flows for heat transport with multiscale perspectives. Section 4 is dedicated to discussion on the motile microorganisms that co-exist in our ecosystem with widespread size, swimming velocities, and unique interacting mechanisms while flourishing in nature. We revisit the planktons' interactions in natural and laboratory settings, especially under the influence of light that can help us to gain more control relevant to thermofluidic behavior. Following which, novel taxis mechanisms and relevant length scales are discussed that necessitate further in-depth understandings of interplay of physical properties stemming from these distinct length scales. The interrelated rheological characteristics and bioconvection influences are briefly discussed afterwards. To revisit the promise of the bioconvective phenomena towards thermal management, we discuss the most relevant experimental observations in Section 5 where collective behavior in bioconvection can impart crucial implications with observed flow substantially more than cell scale flows. Moving forward, we come to that point where we discuss the limitations and challenges this unique transition is about to face and what novel and efficient techniques can be developed while employing such living, synthetic, and bio-inspired components in technologies in Section 6. We ultimately take an attempt to investigate the current endeavors involving combined experimental and theoretical approaches that can be efficient while designing the future bioconvective heat transfer strategies.

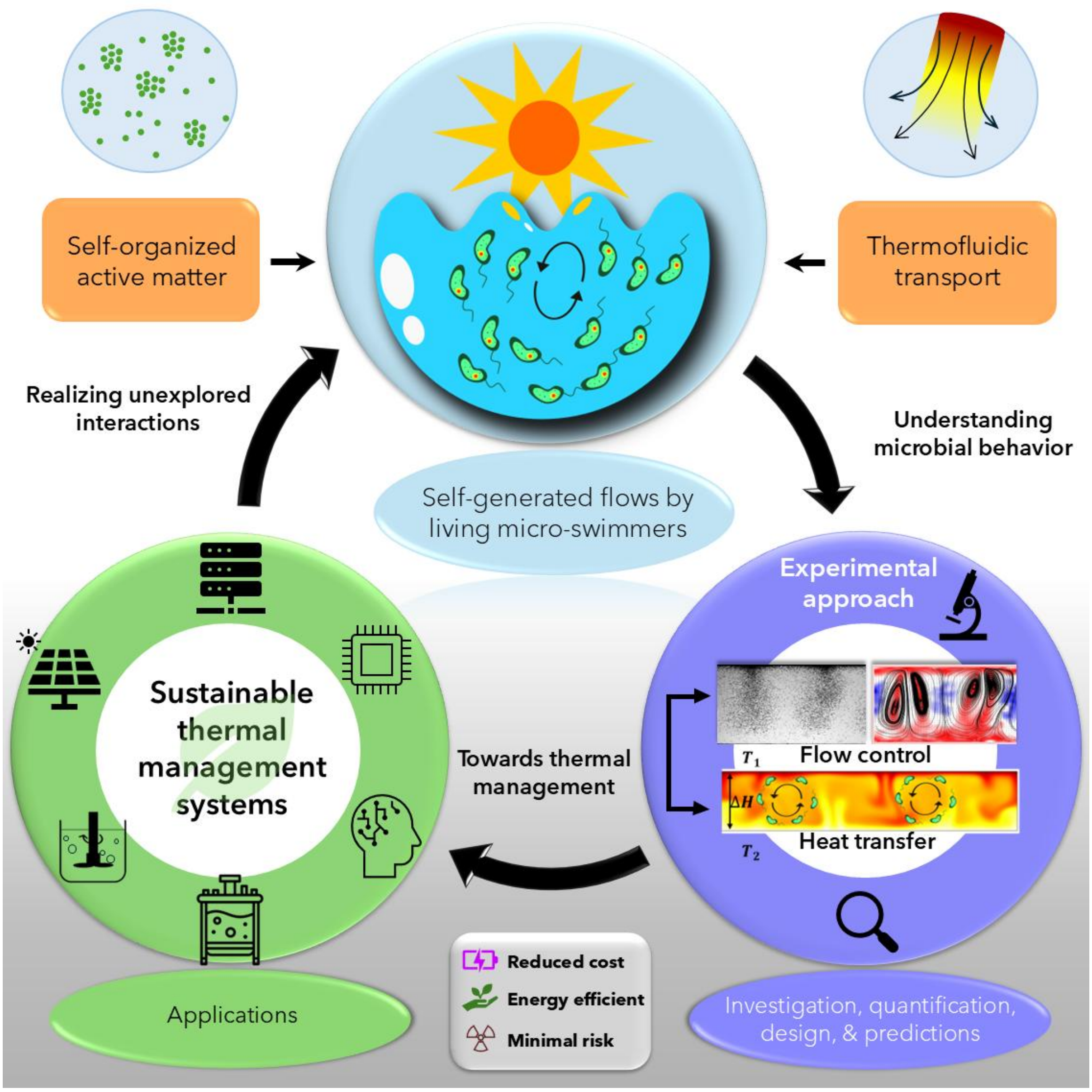

**Fig. 1. Harnessing collective microbial dynamics for sustainable thermal management.** The ability to engineer collective microbial dynamics for thermofluidic transport opens up key possibilities in mitigating largescale energy consumption, typically arising due to cooling of data centres. Projections indicate further rapid growth in energy expenses over the coming decades [17-22,24,26].

## 2. Flow mediated heat transfer

In heat transfer applications, flow mediated thermal management strategies have gained substantial attention due to their effectiveness and applicability in wide varieties of applications [38,63,64]. This encompasses all mechanisms by which fluid motion whether externally imposed, buoyancy-driven, or internally generated by the fluid's constituents, augments heat transport beyond the conductive control [65–67]. To maximize the thermal energy transfer rate from surface to the bulk, novel geometries ranging in multiple order of length scale have been designed, dynamic flow rates have been imposed, and novel working fluids have been introduced [67,68]. The flow mediated heat transfer problem was addressed through the classical triad of natural, forced, and mixed convection, with Nusselt number as the primary figure of merit and the Rayleigh, Reynolds, and Prandtl numbers as the governing parameters. This extensive research is well validated and described in several comprehensive reviews. Overall, the understanding of the fluid properties, geometry, and orientation that influences heat transfer remains central to optimising thermal engineering systems. Heat transfer fluids with constituents such as particles, synthetic organisms, and even swimming organisms have gained more attention in recent years where this inclusion has made it possible to passively modify the effective transport properties of the medium as a whole [14,69,70]. Where for passive particles, the transport properties enhance due to this inclusion with complexities concerning their suspension characteristics, active swimming particles add up additional dynamic intricacies while generating microscale flow distribution that can speed up the heat transfer to large extent. Hence, we motivate ourselves towards these two distinct flow-mediated heat transfer regimes. The first is passive particle driven heat transport in which dispersed inclusions alter the effective thermal conductivity, viscosity, and hence Nusselt number of the carrier fluid without imparting any autonomous motion. The second, and the subject of growing interest with recent advancement in micro-nanofabrication technologies is active organisms and micro-swimmer-driven heat transfer, in which self-propelled entities convert inherent chemical or metabolic energy into directed motion, resulting in coherent convective flows.

### *2.1. Passive particle-driven flow-mediated heat transport*

The concept of suspending finely divided solid particles in a liquid to augment its thermal conductivity predates the modern nanofluid literature by several decades where colloidal particles are suspended with at least one dimension below $100\,\mathrm{nm}$. The metals and metal oxides have thermal conductivities one to three orders of magnitude higher than water, ethylene glycol, or oil, and suspending them as nanoparticles produces an effective medium whose conductivity exceeds that of the base fluids [68,71]. Intriguing mechanisms such as nanolayer formation at the fluid-particle interface, Brownian motion facilitating nanoparticle dispersion and enhancing particle-fluid interactions, and thermophoresis accelerating heat transfer by driving nanoparticles towards cooler regions in a thermal gradient have all been identified as contributions to thermal enhancement in nanofluids, though there is considerable likelihood that they may not be applicable across all nanofluid systems due to the diverse behaviour of different nanomaterials in different base fluids [70,72–77].

The experimental record on nanofluid heat transfer enhancement is extensive but contested. The mechanistic picture is complicated by particle agglomeration, surface contamination, the dependence of enhancement on shear rate and temperature history, and the fundamental difficulty of comparing experiments at equal Reynolds number versus equal pumping power. Measurements of the Nusselt number for water–Cu nanofluid in turbulent regime provided a 30% increase compared with the Dittus–Boelter correlation for 2% volume fraction of nanoparticles [78], results that cannot be explained on the basis of a homogeneous flow model; moreover, the comparison must be performed at the same pumping power rather than the same Reynolds number, since nanofluids require higher average velocity than base fluids to achieve the same Reynolds number due to their increased viscosity. The extension of nanofluid research to porous media has been a particularly active sub-field [70]. In comprehensive reviews of thermohydraulic characteristics of nanofluids in porous media, different aspects including magnetohydrodynamic flows, nanoparticle shape effects, thermophoresis, and hybrid nanofluids were discussed across natural, forced, and mixed convection regimes [70,79,80]. The transition from laminar to turbulent flow represents a threshold above which passive enhancement mechanisms become substantially more effective: in graphene–water nanofluid, the maximum heat transfer enhancement of 36% was observed at a Reynolds number of $3{,}950$ in the transitional regime, with the result indicating that thermophoresis and Brownian motion are more effective heat transfer augmentation mechanisms beyond the laminar regime [81].

However, nanoparticles have no autonomous response to temperature gradients, concentration fields, or local flow conditions; their redistribution is ruled entirely by diffusion and thermophoresis. Hybrid nanofluids, surface-functionalised particles, and carbon-nanotube suspensions represent further elaborations of the same passive paradigm. The inability of passive particles to generate convective velocity fields, to act as internal stirrers where natural instances of swimming organisms often redistributing near and far-field environments motivate the current research endeavours towards the transition to the active paradigm discussed in the following section.

### *2.2. Active (living and non-living) flow-mediated heat transport*

The second category of particle-driven flow-mediated heat transport departs from the passive one due to self-propelling capabilities. This encompasses two distinct families of active agents. The first consists of synthetic micro-swimmers such as Janus particles, catalytic bubble-propelled particles, thermophoretic swimmers, and electro-hydrodynamically driven colloids that convert chemical, optical, or electrical energy into directed locomotion [82,83]. The second one which still seeks more attentions and is our centre of attraction in this article, is living microrganisms such as bacteria, microalgae, and ciliates whose swimming can generate flow patterns and self-organization. In both of these cases, the locomotion is not imposed by an external forcing but generated internally by the agents themselves.

The experimental demonstration that self-propulsion can enhance convective heat transfer was provided recently in direct and compelling form [12,84,85]. Micron-scale self-propelled particles that convert chemical energy into autonomous motion were shown to improve the convective heat transfer coefficient in a pool of suspension heated from below, with enhancements associated with self-propulsion most pronounced at low heating powers where the Rayleigh number is approximately $10{,}000$, and the heat transfer coefficient can be significantly high in the self-propelled case compared to the same particles without propulsion [12]. This result is particularly significant because the enhancement is largest precisely where conventional natural convection may become weak, at low $Ra$, below the threshold for vigorous thermally driven rotation, establishing self-propulsion as a mechanism that is complementary to, rather than competitive with, classical convection.

Among synthetic swimmers, the dynamics of active droplets and Janus particles in thermal environments have attracted growing theoretical and experimental interests [86]. Responsive micro-swimmers powered by electro-hydrodynamic flows can adapt their motility via internal reconfiguration in which light absorption induces preferential local heating and triggers structural transitions, leading to an adaptation of the cluster's motility; this coupling between self-propulsion and thermal state demonstrates a design principle for active particles that respond adaptively to their thermal environment. The temperature dependence of micro-swimmer dynamics is non-trivial and, in some cases, counterintuitive. On the contrary, for biological micro-swimmers, the connection between locomotion and heat transfer operates through the mechanism of their single cell level to collective response and often involving macroscopic circulation patterns driven by the collective upswimming in a dense suspension. Bioconvective flows have been shown to play an imperative role in promoting growth within stratified lakes, where biomixing is able to redistribute thermohaline profile over thicknesses of up to several meters [87]. The control of such flows by light, can engineer arbitrary spatiotemporal patterns, which provides a promising tool to determine microbial accumulation and thereby modify and control macroscopic flows within active suspensions [87]. The photo-bioconvection studies established that the spatial structure of bioconvective plumes can be shaped in real time through structured illumination, providing a control strategy for flow-mediated heat transfer that has no analogue in any passive or conventionally forced system [88].

The active matter literature has also extensively addressed the question of enhanced diffusivity and mixing in suspensions of micro-swimmers [89]. Ensembles of biological and artificial micro-swimmers can produce long-range velocity fields with strong non-equilibrium fluctuations, resulting in a dramatic increase in the diffusivity of embedded passive tracers, a mechanism directly relevant to scalar transport and multiscale self-generated flows with thermal mixing.

The distinction between passive and active particle-driven heat transfer is not merely taxonomic. It reflects a fundamental difference in the energetic origin of the convective enhancement. In passive suspensions, any improvement in $Nu$ beyond that of the base fluid must ultimately be paid for by increased pumping power, the improved effective conductivity reduces thermal resistance but does not reduce mechanical resistance, and the two are coupled through the Reynolds number. In active suspensions, by contrast, the mechanical work of convective stirring is supplied by the metabolic or chemical energy stored in the swimmers themselves, not by an external mechanical input. This decoupling of thermal enhancement from pumping power hence can become the central thermodynamic argument for organism-driven thermal management and represents the conceptual bridge between the well-established physics of bioconvection and the emerging engineering application of active heat transfer fluids. It has already been observed that the bioconvection may substantially influence the flow fields, concentration layout, and temperature up to meters, in natural lakes [87]. Laboratory experiments are witnessing its versatility in terms of flow magnitude and orders of length scale, that may allow to regulate heat transfer as portrayed in Fig. 2. This self-propulsion has therefore motivated researchers to go for active synthetic particles and employing them in heat transfer applications with promises towards mitigating external energy inputs [12,82] (as portrayed in Fig. 2e). The following sections of this review develop this argument in quantitative detail, from the theory and experiments concerning active matter flows and bioconvection to the design of experimental-theoretical systems capable of characterising the future thermal management strategies.

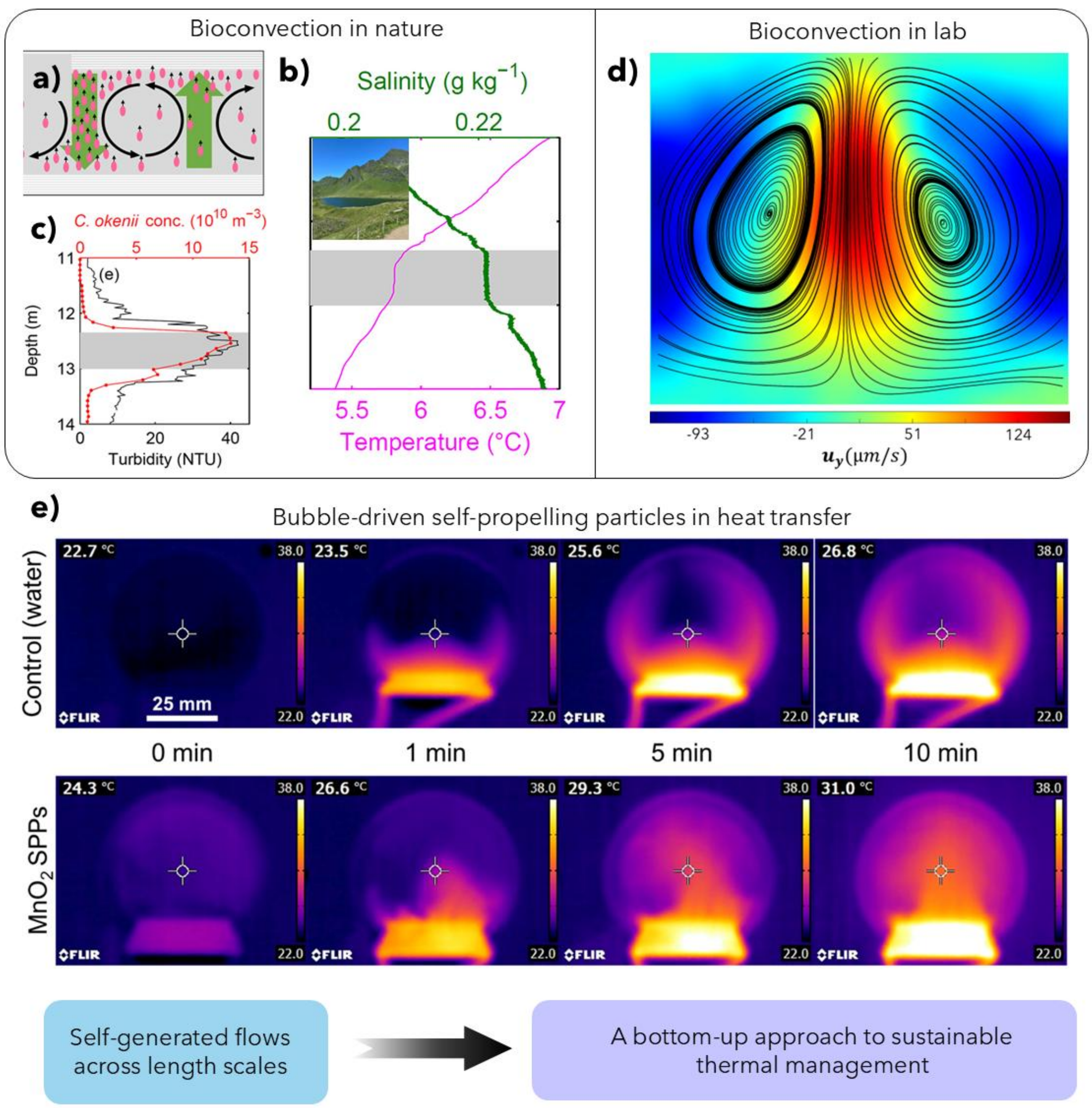

**Fig. 2. A bottom-up approach to sustainable thermal management.** (a) Schematics of bioconvection pattern and circulation observed in a natural lake in Cadango, Switzerland (Credit - Sommer *et al.*, 2017) [87] (https://doi.org/10.1002/2017GL074868). (b) Bioconvection induces thermal transport indicated by the homogenization of the temperature and salinity profile over a meter-scale depth. (c) High accumulation of a motile photosynthetic Sulphur bacterium *Chromatium okenii*, driving large-scale mixing by bioconvection. (d) Particle image velocimetry of a bioconvecting marine alga observed in Lab [90]. (e) Bubble-driven self-propelled nanoparticles for heat transfer reported by Velazquez *et al.* [12] (https://doi.org/10.1016/j.ijheatmasstransfer.2025.127966).

Now that we have discussed flow-mediated heat transport and passive to active suspension, the next question that arises is whether active matter flows can be useful towards future thermofluidic management applications. In the next section, we introduce the bioconvection phenomena first followed by dedicated calculations realizing the heat transfer potential based on the previous observations, thermofluidic property characterization, and applicable correlations of heat transfer metrics. Besides dedicated experiments, theoretical frameworks that can predict thermofluidic performance beforehand will be invaluable while saving time, effort, and money in future experiment designs and device developments. Obviously, active matter flows such as bioconvection involve intricate phenomena stemming from multiple length scales and theoretical frameworks that can predict accurately would not be straightforward and easy to implement, and hence, we discuss a systematic approach for model formulation to capture physics from macro to microscale with primary focus on collective bioconvection instances.

### 3. Bioconvection: from theoretical insights to engineering solutions

Bioconvection emerges when motile microorganisms create density gradients that destabilize suspensions while driving spontaneous convective flows. This is prominent in bioconvection that, collectively, microorganisms may achieve more than they can do as individuals. Collective scale Reynolds number is observed to be hundred times more than the cell scale Reynolds number where self-organised circulatory flow fields are observed and this indicates that the hydrodynamic field with self-propelling motion can surely be used for sustainable heat transport mechanism. Next, we would discuss about the onset of bioconvection and what are the crucial parameters that may influence the phenomena substantially. Following this, we would discuss briefly about the key taxis mechanisms, scales, and non-dimensional numbers that make analysis tractable.

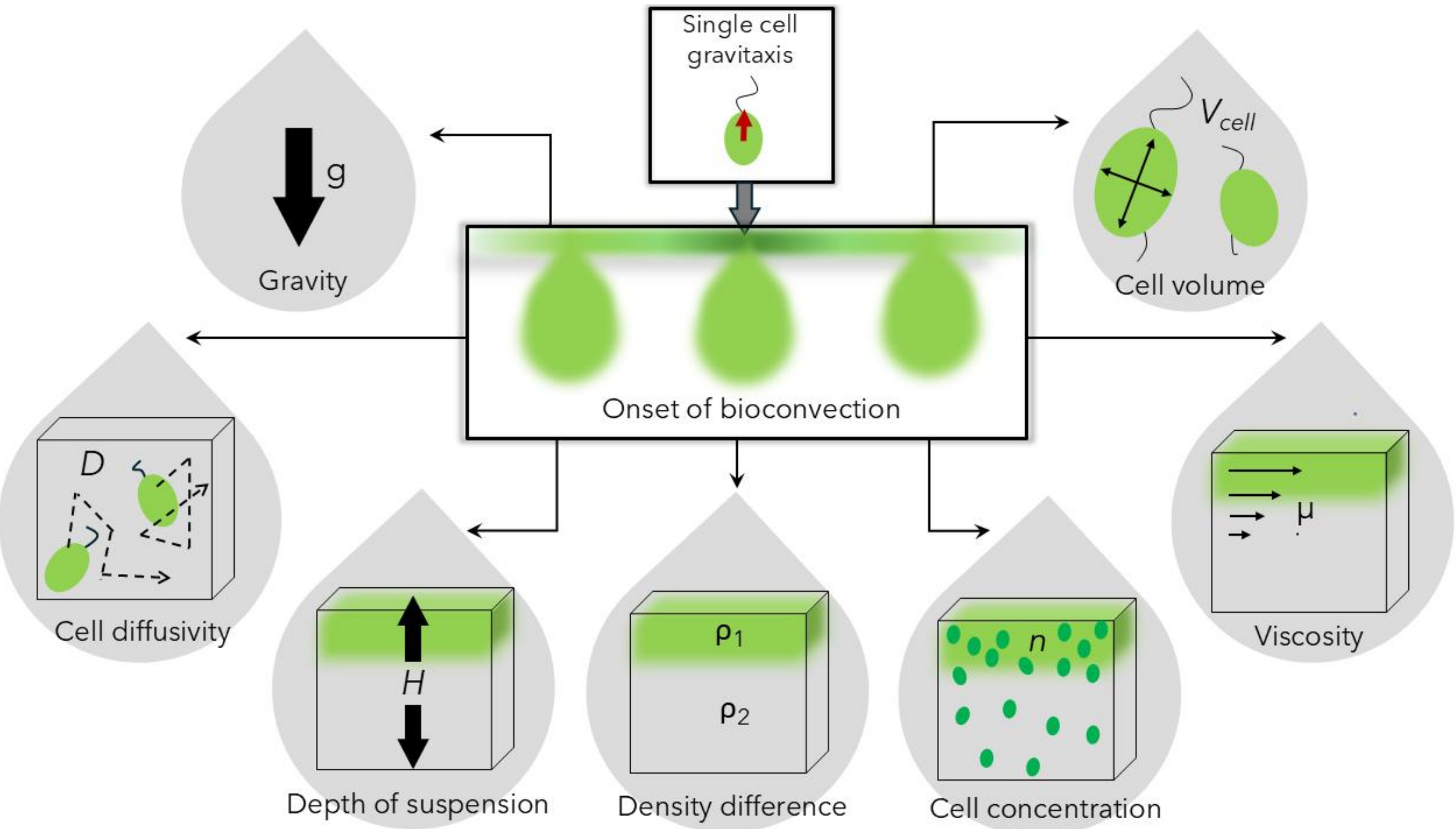

**Fig. 3. Biophysical factors governing bioconvective onset.** The onset of bioconvection is depicted where the phenomena is largely dependent on various properties such as gravity ($g$), cell volume ($V_{cell}$), diffusivity ($D$), depth of suspension ($L_{ch} = H$), density difference ($\Delta\rho$), concentration ($n$), and rheological properties (such as viscosity, $\mu$).

### 3.1. The onset of bioconvection

The onset of bioconvection is governed by an interplay between cellular behavior and environmental conditions [46,48,88]. Although, cell concentration is a primary control parameter: increasing the number of cells enhances the density excess of the upper layer and thereby strengthens the gravitational instability, however, concentration alone is not sufficient. Persistent upward migration is also required to maintain a dense cell-rich layer at the top. If the effective diffusivity of the cell population is high, directional accumulation is weakened and the top-heavy layer can be dispersed, even at high cell concentrations. Cellular swimming behaviour and gravitactic reorientation therefore influence the effective diffusivity of cells and, together with concentration, determine the formation and stability of the upper dense layer.

Additionally, the depth of the suspension can influence pattern formation [91,92], particularly in shallow chambers where wall interactions become important [48]. The excess density of individual cells relative to the fluid further determines the magnitude of the density contrast between the accumulated upper layer and the bulk suspension. In contrast, fluid viscosity acts to stabilize the system by resisting the flow generated by gravitational sinking; consequently, even a dense upper layer may fail to produce visible plumes in a sufficiently viscous suspension. The combined effects of cell concentration, cell excess density, motility, gravitactic reorientation, diffusivity, fluid viscosity, and confinement can be summarized (shown in Fig. 3) through a critical threshold parameter known as the bioconvective Rayleigh number, defined as,

$$Ra = \frac{\left(n\,\Delta\rho\,V_{cell}\mathrm{g}\,{L_{ch}}^{3}\right)}{(\mu\,D)}$$

where $\Delta\rho = \rho_1 - \rho_2$, is the density difference between top and the bulk layer, $V_{cell}$ is the volume of the cell, $n$ is the concentration of the cell, $L_{ch}$ is the characteristic length, $\mu$ is the viscosity, $\mathrm{g}$ is the gravitational acceleration, and $D$ is the diffusivity.

### 3.2. Characterising the potential of microbial induced heat transport

Now, towards exploring the potential of bioconvection for the heat transport applications leveraging the knowledge from active matter physics, we would resort to relevant non-dimensional numbers and performance characterisation first to gather some overview for the thermofluidic behaviour it may offer. Therefore, we enlist first the heat transfer metrices that we need to characterise for realising the thermal performance of bioconvection or other active matter driven heat transfer.

Among the relevant heat transfer metrics, the first comes the heat transfer coefficient to quantify how effectively heat can be removed from a surface and the ratio of heat transfer coefficient of bioconvective or bio-organism flow driven heat transfer to the passive system. Nusselt number ($Nu$) serves as the primary quantifying parameter for relative influence of heat transfer compared to the convective part. Concurrently, the thermal resistance provides a practical assessment of the ability of bioconvective suspensions to dissipate heat from a source. Reynolds number comes naturally into the discussion as the inertia to viscous comparison within the flow system and Péclet number due to its relevance in characterizing the strength of the advective heat transport arising from flow structure-driven mixing compared to molecular thermal diffusion. Establishing quantitative relationships among these metrics would provide a unified framework for understanding how microorganism-induced flow structures modify thermal boundary layers, augment heat transport, and ultimately determine the feasibility of bioconvection-driven thermal management systems. Next, we take an attempt to characterise and quantify to what extent microbial flow mediated heat transfer specially bioconvection can be self-sufficient or energy efficient in terms of heat transfer. Considering aqueous nutrient media standard in bioconvection investigations we may assume the following values for relevant properties.

**Box 1.** *Relevant heat transfer metrics towards characterising thermofluidic performance*.

| Heat transfer metrics | Expression | Physical significance |
|---|---|---|
| Heat transfer coefficient ($h, \mathrm{kW/m^2}$) | $h = \frac{\dot{Q}}{T_a - T_b}$ | Effectiveness of heat removal |
| Nusselt number ($Nu$) | $\frac{hL_{ch}}{k}$ | Ratio of heat transfer to conduction |
| Thermal Resistance ($R_{th}$) | $\frac{\Delta T}{Q}$ | Resistance offered to heat flow |
| Reynolds number ($Re$) | $\frac{\rho u L_{ch}}{\mu}$ | Ratio of inertia to viscous forces |
| Thermal diffusivity ($\alpha$) | $\frac{k}{\rho c_P}$ | Competition between heat conduction and thermal storage |
| Péclet Number ($Pe$) | $\frac{uL_{ch}}{\alpha}$ | Ratio of convective to diffusive transport |

where $\dot{Q}$ is the heat transfer rate, $T_a$ and $T_a$ are the temperature at surface $a$ and $b$, $Q$ is the transferred heat, $c_P$ is the specific heat, and $u$ is the velocity. Thermal conductivity ($k$) = $0.60\ \mathrm{W\,m^{-1}K^{-1}}$, density (ρ)= $998\ \mathrm{kgm^{-3}}$, specific heat ($c_p$)= $4180\ \mathrm{Jkg^{-1}K^{-1}}$, $\alpha = 1.43 \times 10^{-7}\mathrm{m^2s^{-1}}$.

Previous experimental observations confirm that bioconvection can generate rms velocities of $0.1 - 1\ \mathrm{mm/s}$ when the cell swimming velocity may lie within $50 - 300\ \mathrm{\mu m/s}$ which is sufficient to drive laminar and even transitional convection with $Re = \frac{\rho V_S L_{ch}}{\mu} \cong 10 - 100$, $Pe = \frac{V_S L_{ch}}{\alpha} = 0.3 - 2.1$, and ultimately resulting in $Nu = 2 - 10$ across typical lab scales [93–95]. Next characteristic thermofluidic property that we would mention here is the effective diffusivity enhancement in bioconvection. The active mixing generates a higher effective diffusivity surpassing passive diffusion of heat up to ten orders. However, the bioconvection along with other nanoparticles may increase the effective thermal conductivity ($k_{eff} = 0.6 - 1.5\ \mathrm{Wm^{-1}K^{-1}}$) [96]. Considering higher dimensions comparable to larger heat management devices ($H = 10 - 50\ \mathrm{cm}$), the self-pumped flow may deliver heat fluxes of $1 - 10\ \mathrm{kW/m^2K}$ [12,84] at metabolic power densities than mechanical pumping. For a $10\ \mathrm{kW}$ data center rack, this can save up to $10{,}000$ € annually estimated per rack at current electricity prices, positioning bioconvection as a transformative technology for sustainable cooling in moderate heat-flux applications. If we consider the system geometry similar to a conventional data centre cold plates where the channel height may vary at the range of $0.5 - 2\ \mathrm{mm}$ and channel length in the range of $100 - 300\ \mathrm{mm}$, we did calculations for heat flux, Nusselt number, Péclet number, Reynolds number and pumping power, and we present the estimation in Box 2.

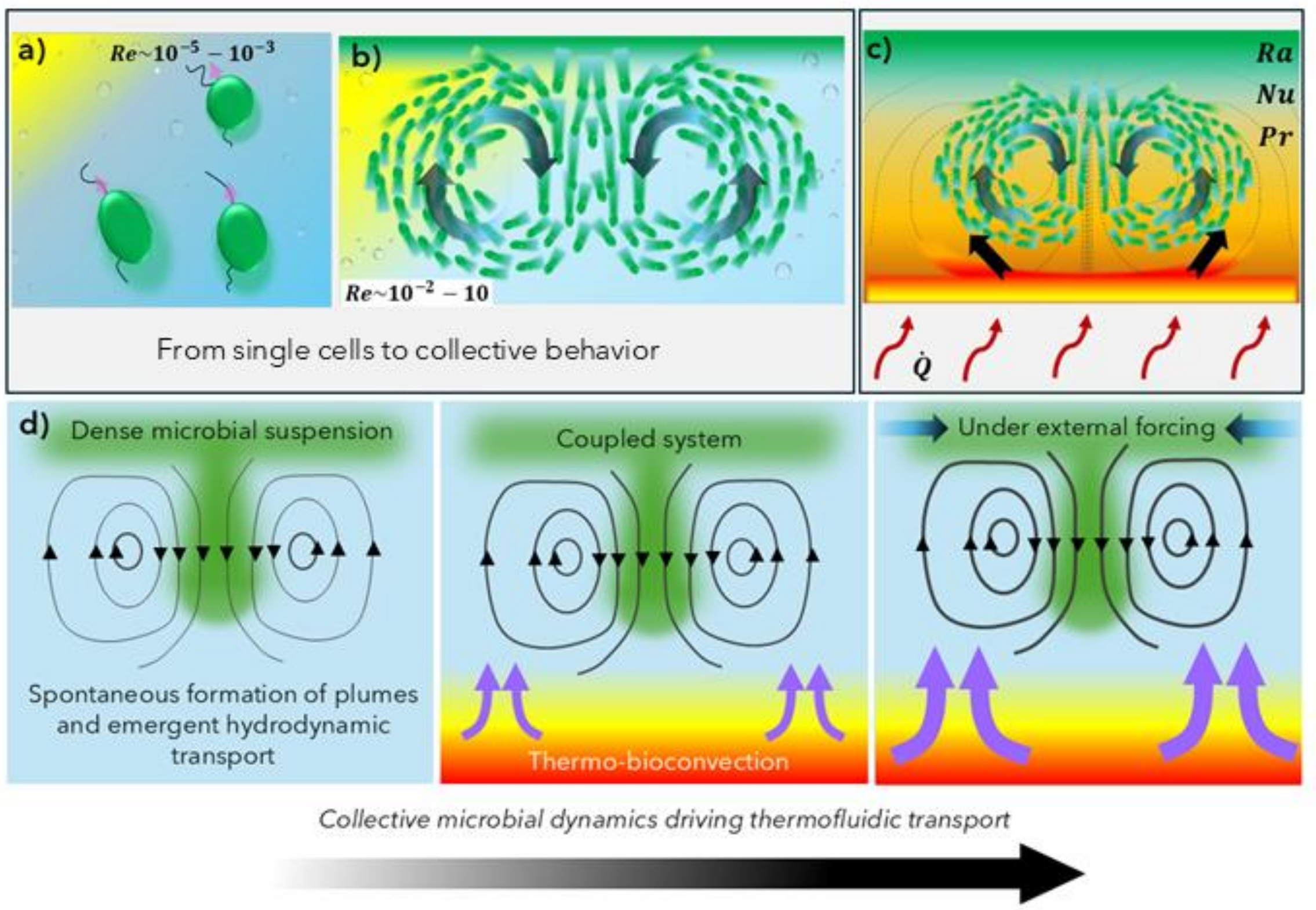

**Fig. 4. Active thermofluidic transport due to bioconvection.** (a) At single cell level, the Reynolds number is observed to be smaller than the Reynolds number of collective scale active flows by two orders of magnitude (b) the collective behavior resulting in self-generated plumes may have the potential towards future sustainable thermal management devices while redistributing heat. (c) Bio-inspired thermofluidic transport can be characterized by the heat transfer metrics such as Nusselt number, Prandtl number, Rayleigh number, and the temperature fields/gradients. (d) The schematic representation of enhanced mixing in coupled systems where the flow circulation by natural convection gets aided due to bioconvection. In the presence of external forcing, this mixing can be further enhanced.

Fig. 4 portrays the schematic of microorganism taking part in collective flows that may involve much higher Reynolds number compared to observed ranges at the cell scale (Fig. 4a-b). This self-generated flow structure can work as a driving mechanism with potential to enhance heat transfer beyond the diffusion limit (as shown in Fig. 4c). The disruption of the thermal boundary layer via the motile microorganisms can be highly beneficial to increase the effectiveness of the heat transfer strategies (Fig. 4d). We delve deeper into the organism specific biophysical parameters extracted from the primary literature to compute effective Nusselt number (shown in equation 1 and 2), heat transfer coefficient, and thermal characteristics with valid assumptions and established correlations. Unlike the fundamental formulation or Nusselt number, we consider classical correlations from literature to revisit the thermal characteristics for an enclosed horizontal fluid layer heated from below this time [97,98].

$$Nu = 0.069 \times Ra^{1/3} \times Pr^{0.074}, \text{ for } Ra \gg 1500 \quad (1)$$

$$Nu = 1 + 1.44 \times max\left(0,1 - \frac{1708}{Ra}\right) + max(0, \left(\frac{Ra}{5830}\right)^{\frac{1}{3}} - 1), \text{ for } Ra \sim 1500 \quad (2)$$

Under the dilute approximations ($\varphi \ll 1$ $and$ $\frac{\Delta\rho}{\rho} < 0.1$, where φ is the cell volume fraction), the thermal and bioconvective buoyancy forces enter the Boussinesq momentum equation additively and we consider an effective Rayleigh number as, $Ra_{eff} = Ra_T + Ra_b$ as a first-order estimate. $Ra_T$ and $Ra_b$ are respectively the thermal and bioconvective Rayleigh numbers (discussed in more details in section 4.4). However, the Nusselt number may vary depending more on one particular effect when one diffusivity governs the whole instability and hence, we have additionally calculated the ratio of bioconvective to thermal Rayleigh numbers. For the cases considered here, it is observed that the ratio of the bioconvective to thermal Rayleigh number is lying within the range of $0.1$ to $10$ and hence we are sticking to the first order approximation as discussed earlier. Considering two different species with the chamber depth ($H$) varying in the range of $5$ mm to $20$ mm, we calculate approximate heat transfer metrics which are $Nu$, $h$, and possible heat flux this time. Table 1 enlists the relevant parameters considered for the species in the calculation taken from respective previous research [47,48,58,60,93,95,99,100].

**BOX 2:** *A projective quantification of the bioconvective heat transfer based on measurable traits in experiments.*

| Metric | Pure conduction | Bioconvection | Forced liquid cooling | Hybrid bioconvection |
|---|---|---|---|---|
| Nusselt number ($Nu$) | $1$ | $1.2 - 9$ | $10 - 20$ | $12 - 25$ |
| Péclet Number ($Pe$) | $0$ | $0.3 - 2.2$ | $100 - 500$ | $10 - 100$ |
| Reynolds number ($Re$) | $0$ | $10 - 100$ | $1000 - 2000$ | $200 - 500$ |
| Pumping Power ($\mathrm{W/cm^2}$) | $0$ | $< 0.01$ | $0.2$ | $0.04$ |

For the base-fluid thermophysical properties, we are considering sea water for *H. akashiwo* and fresh water for other species. Now, incorporating the relations from equations (1) and (2) and calculating the Rayleigh numbers, Nusselt number, and $h$ for *H. akashiwo*, we obtain $Nu$ and heat transfer coefficient, $h$. Effective thermal conductivity is assumed to vary by $10 - 20\%$ compared to the base fluids. Similarly, considering properties for *C. reinhardtii*, we analyzed the heat metrics, and we observe that the Nusselt number ranges are mostly observed to match our previous calculations while ranging between 2-10 with slightly larger values for *H. akashiwo* compared to *C.reinhardtii. T*he heat transfer coefficient is observed to vary in the range of $150 - 300\,\mathrm{W/m^2K}$. Though, *H. akashiwo* is observed to offer larger heat transfer coefficient compared to the other, the dedicated experiment can only reveal the true species specific thermofluidic behavior.

***Table 1.*** *Microorganism properties considered in the calculation.*

| **Organism** | $n \times 10^5\ (\mathrm{ml^{-1}})$ | $V_{cell} \times 10^{-16}\ (\mathrm{m^3})$ | $V_s(\mathrm{\mu m/s})$ | $\Delta\rho/\rho$ |
|---|---|---|---|---|
| *Heterosigma akashiwo* | $\sim 1$ | $\sim 2.5$ | $\sim 65$ | 0.03 to 0.05 |
| *Chlamydomonas reinhardtii* | $\sim 1$ | $\sim 5$ | $\sim 110$ | 0.05 to 0.06 |

From our calculations, we observe that the heat transfer performance may largely vary depending on several factors relating to cell properties such as size, volume, and density to influencing properties of the base fluids and chamber size. Furthermore, cell activity additionally may vary depending on the environmental conditions and other constituents if present. The full picture can only be disclosed if dedicated experiments and analysis are performed on different species in controlled environment with precise variation of influencing parameters. Due to larger cell size, $Ra_b$ may become higher for *H. akashiwo* compared to *C. reinhardtii*. However, in actual situations, the living motile cells can show additional non-intuitive dynamic features which are yet to be revealed in future. Future analysis involving more theoretical details may unveil such non-intuitive interactions in bioconvective heat transfer systems. Furthermore, instances like these may result in double-diffusive convection (DDC) stemming from the contributions to fluid density stratifications due to cell distribution and thermal fields with competitive stabilizing and destabilizing effects. The phenomenon of double-diffusive convection was first identified in the ocean by Stern [101] where it was observed that a fluid stable to either gradient alone can become unstable once both are present, owing to the differential diffusion of heat and salt. Starting from Turner's review [102] establishing density-ratio framework, Hupper & Turner [103] extended into a comprehensive account of finger growth, flux laws, and staircase formations. As in coupled thermal-bioconvective system, the temperature field and cell-accumulation both contribute to the buoyancy, and they diffuse at different rates, there can be complex layering due to DDC. From the DDC theory predictions, when the buoyancy ratio between the two fields is of order of unity and the diffusivities differ substantially, the system can organize into layers with transport rates that may depart from what a simple addition of Rayleigh numbers would suggest.

### *3.3. Detailed theoretical frameworks*

While formulating the governing equations, we may first look into the ranges of the relevant non-dimensional numbers to get an overview of the nature of the flow-field in collective active matter flows and bioconvection. The individual cell Reynolds number is in the order of $10^{-3}$ where the bioconvection Reynolds number may reach up to 10 to even 100 [48]. The model previously used is generally based on the incompressible Navier-Stokes equations with an additional buoyancy term subjected to a Boussinesq approximation representing the influence of the difference in density between the cell and the fluid medium ($\Delta\rho$), mean volume ($\vartheta$), and cell concentrations ($n(x,t)$) [104,105]. Consequently, the Navier-Stokes equation now becomes,

$$\rho \frac{D\boldsymbol{u}}{Dt} = -\nabla P_e + n\vartheta\Delta\rho g + \nabla \cdot \Sigma, \quad \nabla \cdot \boldsymbol{u} = 0 \tag{3}$$

Here, $\boldsymbol{u}(x,t)$ is the velocity of the suspension, $n(x,t)$ is the cell concentration, $P_e$ is the excess pressure, and $\rho$ and $g$ are the fluid density and gravitational acceleration respectively. The cell-induced stress $\Sigma$ was first considered by Pedley and Kessler in their formulation [105] which has three components namely swimming-induced stresslets, Batchelor tresses, and rotary particle diffusion associated stress. However, the last two components can be neglected and for the first term, considering the dominating leading-order stress for the aggregate swimming stroke of the cells, the stress term can be expressed as,

$$\nabla \cdot \Sigma = \mu \nabla^2 \boldsymbol{u} + \nabla \cdot \Sigma^{(p)} \tag{4}$$

Here, $\mu$ is the viscosity of the fluid medium. Since the timescale for bioconvection is shorter than for cell growth and the cell conservation equation can be written as,

$$\frac{\partial n}{\partial t} = -\nabla \cdot \mathrm{J}, \text{ with flux } \mathrm{J} = [n(\boldsymbol{u} + \boldsymbol{V_s}) - D \cdot \nabla n] \tag{5}$$

The formulation was proposed and used in Pedley *et al.* [45] and Pedley & Kessler [105], where the flux contains three terms: one for the advection by flow, one for the drift relative to the flow with mean swimming velocity, and one for swimming diffusion with diffusion tensor $D(x,t)$.

By establishing torque balance, the orientation $\mathbf{p}$ of a spheroidal swimming cell was expressed deterministically by Pedley & Kessler [106,107].

$$\dot{\mathbf{p}} = \frac{1}{2B}[\boldsymbol{\kappa} - (\boldsymbol{\kappa} \cdot \mathbf{p})\mathbf{p}] + \alpha_0[\mathrm{E} \cdot \mathbf{p} - \mathbf{pp} \cdot \mathrm{E} \cdot \mathbf{p}] \tag{6}$$

Appropriate forms for $\boldsymbol{V_s}(x,t)$ and $D(x,t)$ vary for each taxis mechanism and for better interpretation, under certain conditions, Eqn. (6) can be established directly by approximating a Smoluchowski equation considering the probability that a cell may have a particular orientation and position [108,109]. However, at the early phase, gyrotaxis were described as deterministic [106] (equation 6) and it was realized that such considerations were inconsistent for swimming orientation following which a Fokker-Planck equation for probability density of swimming orientation was employed later [105]. To approximate the cell swimming diffusion tensor, Pedley & Kessler [105] proposed diffusion approximation expressed as, $D = {V_s}^2 \tau var(\mathbf{p})$, where $V_s$ is the mean swimming speed. From there, research has implemented different methods like Galerkin method to expand the Fokker-Planck equation in spherical harmonics to find mean orientation [105] and diffusion tensor and Taylor dispersion theory for biased swimming [108]. Among the models that have been developed from time to time, photokinesis-Gyrotaxis, Light-dependent Centre-of-Mass Offset, and Reactive Phototactive Torque [107,110,111] are the three that have remained highly celebrated while considering different aspects of cell swimming behavior while incorporating external stimuli effects and they have further modified to better interpret the bioconvective behavior [111]. To better interpret the cell level dynamic reorientation, we would briefly discuss the force balance equations and reorientation in the section following the energy equation described next.

In order to describe the thermal transport in the bioconvective or bio-organism suspension, one would additionally use advection-diffusion equation, in which the advective term is governed by the microorganism-induced velocity field and the diffusive term accounts for conductive heat spreading through the effective suspensive medium. Consequently, the governing equation can be written as,

$$\rho c_P \left( \frac{\partial T}{\partial t} + \boldsymbol{u} \cdot \nabla T \right) = \nabla \cdot (k \nabla T) + Q_s \qquad (7)$$

where, $T$ is the temperature and $Q_s$ is the heat source within the system if present. The thermal conductivity, diffusivity, and even viscosity can be calibrated with direct experimental observations with varying concentrations and other environmental conditions influencing cell accumulations. The coupled system of equations (3-6) and 7 can be solved to obtain the heat redistributions and heat transfer metrics described in section 3.2. With additional light-driven motility, the swimming velocity ($V_s$) can vary dynamically with concentration, temperature and light intensity. Despite substantial advances in realizing the hydrodynamic instabilities, pattern formation, and collective transport phenomena associated with microbial self-organized flows, its implications for heat transport remain largely unexplored and hence, energy equation has received comparatively little attention in this regard. Concomitantly, extending conventional bioconvection models to include rigorous energy transport analysis may provide fundamental basis for evaluating such microbe-mediated flows such as bioconvection as a self-sustained, biologically driven mechanism for next-generation thermal management and energy transport applications.

### *3.4 Single cell biomechanics of gravitactic reorientation*

Starting from resistive force theory [112] to extended computational approach, involving regularized stokeslets [113] have been useful to investigate gyrotaxis for biflagellate. Cell reorientation mechanisms were explored associated with bottom-heaviness and sedimentation. Moreover, if we focus more on the dynamics of the individual cells and their effects on the bulk flow, first we should get an idea of the size and shape of the microbes we are aiming to investigate. The focus here mostly is on such genera those have the diameter in the range of 5 to 100 $\mu\text{m}$. For such planktons, the swimming velocity ($V_S$) range up to some 100s of microns per second. Now, in general, an individual cell while translating and rotating within a fluid medium experiences external body force, external couple, and intrinsic swimming motions [114]. Considering all conditions, the most relevant body force is due to the gravity and density difference with the fluid medium ($\boldsymbol{F} = \text{g}\Delta\rho/\rho$). The external couple may lead to a translational motion for asymmetric or helical shaped organisms [114]. Swimming organisms display a wide spectrum of locomotory strategies, predominantly relying on the coordinated beating of cilia or flagella to generate propulsive viscous thrust that counteracts hydrodynamic drag on the body, as well as viscous torques that must be balanced by internal and externally induced torques [115,116]. Due to the lower values of the Reynolds number, we may safely ignore the inertial effects in the theoretical considerations while incorporating locomotion mostly governed by the viscous force and hence Stokes-flow approximation has provided satisfactory results in substantial cases [117,118]. As a consequence of their periodic shape deformations, a cell attains a characteristic swimming velocity, $V_S$, relative to the surrounding fluid. Considering a body revolving and swimming in a fluid flow, the translational and rotational equations of motion have been decoupled [41,119–122]. Along the major and the minor axis, the equations can be written as,

$$F_P \sin \emptyset = F_D \sin \theta \quad (8)$$

$$F_P \cos \emptyset - F_D \cos \theta = (\rho_{cell} - \rho_{fluid}) V_{cell} g \quad (9)$$

Here, $F_P$ is the propulsive force due to the flagella beating acting along the long axis with an angle $\emptyset$ to the gravity; $F_D$ is the drag force through the center of the hydrodynamic stress $C_H$, directed opposite to the cell swimming velocity, with an angle $\theta$ relative to the vertical. $V_{cell}$, $\rho_{cell}$ and $\rho_{fluid}$ are volume of the cell, cell density, and fluid density respectively.

The drag force, $F_D$ on the moving organism in fluid with viscosity, μ at velocity $v$ can be expressed as:

$$F_D = F_{D\parallel} \cos \alpha_b + F_{D\perp} \sin \alpha_b \quad (10)$$

$F_{D\parallel}$ and $F_{D\perp}$ are the drag forces along and perpendicular to the direction of the major axis and $\alpha_b$ is the angle between the body axis and the direction of swimming. The torques acting at the center of buoyancy can be expressed as:

$$T_V = T_H + T_W \quad (11)$$

where $T_H$ is the torque generated by the drag force $F_D$, $T_W$ is the generated torque by the self-weight of the cell ($W = V_{cell} \rho_{cell} g$). As the propulsion force does not generate any torque about $C_B$, the net torque balance equation can be written as,

$$L_H F_D \sin(\theta - \emptyset) - W[\sin(\emptyset - \arctan(L_{Nb}/L_{Na}))\, L_W = R\omega\mu] \quad (12)$$

$\arctan(L_{Nb}/L_{Na})$ is the contribution to the gravitational torque coming from the offset $L_{Nb}$ of the nucleus within the equatorial plane; and $R\omega\mu$ is the net viscous torque ($R$ is the coefficient of resistance of the body to rotation and ω is the angular velocity of the cell). The length-scale $L_H$ is the offset between the center of buoyancy and the hydrodynamic stress center, $C_H$ and the position of $C_H$ can be obtained by solving the Navier-Stokes equations around the cell body. More details on the methods to derive and solve the expressions for rotational torque, sedimentation velocity, rotation rate, and coefficient of resistance have been described in previous articles [41,47,123,124].

In recent times, a considerable effort has been invested in theoretical investigation of the heat transfer enhancement that active microbial behavior can offer when used with the nanofluids [77]. Numerical schemes such as Runge-Kutta-Fehlberg method (RKF-45) [125] or bvp4c [126–129] have mostly been used to transform the system of constituent partial differential equations into ordinary differential equations to solve for the temperature field [125,129,130], entropy generations [125], concentration field [130], and consequently relevant non-dimensional numbers (such as Nusselt numbers or Sherwood numbers [127]) under the influence of Newtonian and non-Newtonian rheology [129,130] and external fields [128] in conventional theoretical simplified geometrical configurations. These existing bioconvective heat-transfer studies shine lights on the potential of this in near future despite the fact that there are still unknown facts and non-trivial behavioral dynamics that can only be realized through dedicated experiments integrated with detailed 3D multiscale simulations. These previously mentioned existing models offer to solve highly reduced, 1D or similarity-transformed boundary-layer formulations with generic motile-microorganism models. These investigations are obviously crucial to explore parametric trends, but more specific behavioral traits are to be realized to overcome the knowledge gap relating to the single cell to collective scale behavior and real-life geometrical considerations with multiscale attributes.

### *3.5. Molecular modelling and multiscale approaches*

If we penetrate deeper towards the cell level, the interaction of the cell walls with water, ion, and other materials is to be investigated by molecular approaches such as molecular dynamics [131–136], fast simulation methods [137], Brownian particle models [138], or Monte Carlo simulations. Cell level components such as flagellar motors, photoreceptors, and thermotactic ion channels go through specific conformational changes that drive taxis embedded in complex aqueous environment of water, ions ($\mathrm{Na}^{+}, \mathrm{K}^{+}, \mathrm{Ca}^{+}\ etc.$) and metabolites (ATP, c-di-GMP) [138,139]. Hata *et al.* [140] used MD simulation to investigate the influence of high pressure on the protein binding to the flagellar motor and bacterial chemotaxis. They observed that high hydrostatic pressure increases water density at the CheY-FliM interface of the *E. coli* flagellar motor which disrupts salt bridges and inhibits torque transmission for motor reversal. Molecular dynamics can additionally quantify binding kinetics, torque generations, and polymorphism of flagellar filaments. The challenges in modelling primarily are cell size and unpredictable organism behavior. Possibles approaches are to investigate the interaction of the respective protein structure or to develop coarse-grain modelling [141,142] dedicated to bioconvective microorganisms. In that case, force field development for coarse-grain molecular dynamics is the first thing that is necessitated and experimental observations at the cell or molecular level would be a must needed information. In the coming future, a substantial amount of research is to be done worldwide to shape the full-fledged application of bioconvection in versatile technological fields.

As discussed in previous sections, bioconvection naturally involves phenomena from multiple scales and multiscale approaches would obviously contribute to the future research and technologies. Multiscale approach includes the amplification of smaller scale attributes contributing to the dynamic behavior at the collective scales [143–145]. In bioconvection, the interaction of individual cells with the flow medium and its impact on flow rheology may stem from the molecular scale behavior involving cell structure and particular protein-solvent interactions that influence the continuum properties of the cell-fluid interactions. The insights from cell-fluid interactions and the thermophysical properties of microorganisms and their surrounding microenvironment may bridge the gap between the actual diffusivity with in-depth realization of interfacial thermal resistance, near-surface fluid structuring, temperature-induced modifications in membrane dynamics, and local energy transport mechanism at the vicinity of swimming cells. By coupling the molecular scale interfacial effects and its influence on the continuum level dynamic property distributions, it would serve the purpose towards accurately predicting the bioconvection behavior in terms of physicochemical thermofluidic features.

Following this discussion about theoretical and numerical approaches and a projection towards thermal management applications, we get back to the relevant facts about microbes and swimming microorganisms, their coexistence in nature, and light interactions followed by experimental observations that show promises towards future applications using biohybrid components.

**4. Microbes in nature**

The previous research on motile microbes while showing self-organised flows hints us that they have the potential to be used as a potential candidate in flow manipulation and eventually heat transfer or other flow related technologies. According to the size and relevant length-scale dependence, planktons can be largely classified as bacteria/nanoplankton $(1-5\ \mu\text{m})$, microplankton $(10-100\ \mu\text{m})$ and zooplankton $(0.3-5\ \text{mm})$. Reported swimming speeds span $\sim 1-40\ \mu\text{m}/s$ for bacteria, $0.05-0.5\ \text{mm}/s$ for flagellates and up to $\sim 7\ \text{mm}/s$ for ciliates, and $1-4\ \text{mm}/s$ for larvae with copepod escape jumps reaching $80-200\ \text{mm}/s$ [61,146–148]. At the smallest scales, dense suspensions of nanoplanktons generate mesoscale vortices and even active turbulence [149]; microplankton form gyrotactic plumes and overturning convection patterns; and zooplankton produce jet- and dipole-like flow fields with vorticity extending several body lengths. The nanoplanktons for their small sizes moves with translation characteristics involving Brownian motion effects. One form of collective-scale dynamics emerges through coupled cell-cell and cell-fluid interactions, leading to the formation of spiral vortices. In thin confinements, when the chamber radius falls below a critical value, the suspension can spontaneously organize into a steady single-vortex state surrounded by a counter-rotating boundary layer of cells, and they adopt a spiral orientation within the vortex [137,150]. On the contrary, the taxis behaviour in microplanktons leads to collective motions often depicting consistent flow structures such as bioconvection. Zooplankton, due to their larger sizes and swimming velocities, although may have higher capabilities to induce disturbances in the nearby flow fields, more complex swimming mechanisms, large size, and individuality may add complexities towards immediate applications in technologies. The variability of different planktons in terms of size, induced velocity, their mechanisms, types of flow regimes, and collective behaviour patterns are shown in Fig. 5. In all cases, they show collective behaviour which is largely different from their individual scale.

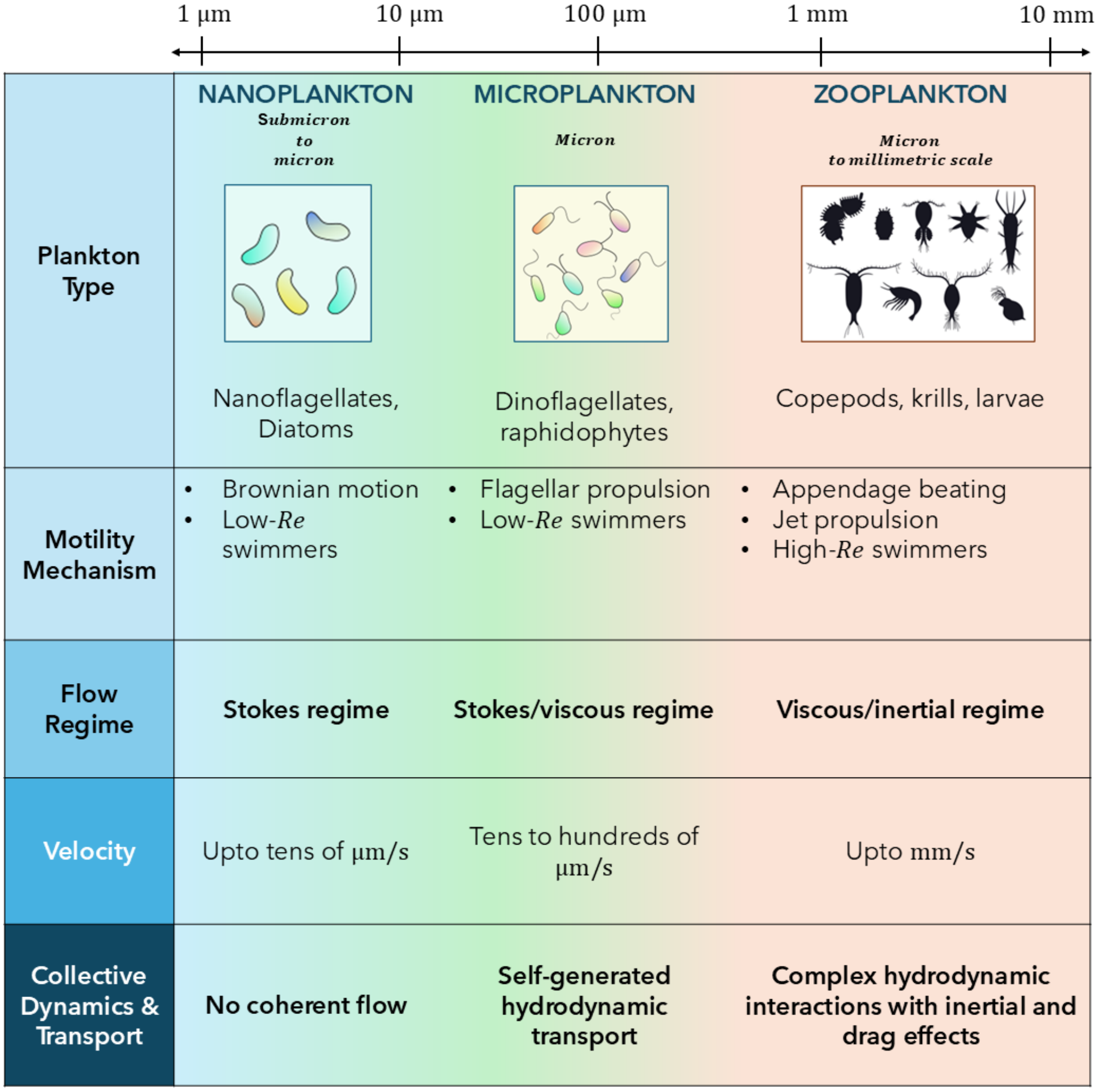

**Fig. 5. Diversity in length scales and motility in microbial active matter.** A range of planktonic species illustrates how individual-scale motility can give rise to collective dynamics, in relation to the cell size and swimming mechanism.

The rest of the responsibilities lies in our hand that concerns to what extent we understand them and manifest their movement patterns corresponding with their size and ranges of the swimming velocity to train, synchronize, and utilise them in the present and upcoming dedicated technological fields. One of the focuses is on the photosynthetic microbes is because phototaxis can be considered as the accurate and attractive one when it comes to the control and steering which remains an important challenge for controlling microbes swimming. However, we have not been totally conservative towards phototaxis, but we have revisited other taxis mechanisms as well in the section **4.2**.

### *4.1. Evolution, lights, and active matter transport*

Marine phytoplankton and cyanobacteria produce a huge share of Earth's oxygen through photosynthesis, supporting higher life, driving key biogeochemical cycles, and transforming carbon, nitrogen, sulfur, and other elements to large extent. They live in close co-existence with plants and animals, forming symbioses (e.g., gut microbiomes, root-associated microbes) that influence health, growth, and behavior [151,152]. At the same time, microbes also cause diseases, harmful algal blooms, and toxin production, reminding us that their “co-existence” with other life is a dynamic balance between beneficial functions and potential harm [153–156]. Motile microorganisms utilize energy to sense, move, and collectively shape their environment and natural selection has favored the strategies that make this activity pay off [157,158]. From an evolutionary standpoint, microbial life represents one of the earliest and most successful strategies for energy transduction and transport in nature [152]. Long before the emergence of complex multicellular organisms, microbes evolved mechanisms to harvest environmental energy, chemical, thermal, or radiative, and convert it into directed motion and metabolic work. In physical terms, these places microbes squarely within the framework of active matter: systems composed of self-driven units that continuously inject energy at the microscale and remain intrinsically out of equilibrium. Unlike passive particles, microbial swimmers generate persistent forces and stresses on their surrounding fluid, leading to emergent collective behavior that cannot be inferred from equilibrium thermodynamics alone. Photosynthetic microbes are particularly notable in this regard, as they directly convert light energy into motility and buoyancy modulation, thereby coupling radiative energy input to hydrodynamic transport [159]. Phytoplankton and their diel vertical migrations are well synchronized with the light conditions throughout the day and night [160], and even by the anthropogenic light activities [161]. Light interacting proteins and receptors such as flavoprotein, phytochromes, and rhodopsins are present that work in different wavelength ranges [162]. Some photosynthetic taxa show directional migration due to bottom heaviness, shape asymmetry, and photosynthetic responses [107,163,164]. Certain motile microorganisms exhibit persistent positive phototaxis while they swim towards maximal illumination. In contrast, some are there which portray a more nuanced phototactic behavior (e.g. *Euglena* and *Chlamydomonas*) being attracted to low-intensity light while actively avoiding high irradiance [165]. In such cases, negative phototaxis can often dominate over negative gravitaxis [166]. In numerous cases, phototactic steering is intrinsically linked to the rotational or helical motion accompanying flagellar propulsion [167]. The adaptive locomotory responses are produced from coupled action of physical and physiological orientation mechanisms. A centrally positioned photoreceptor experiences cyclic shading by an optically opaque organelle, the stigma or eyespot, located near the cell periphery. Temporally resolved photic signals generated from exposure to intermittent light, are transduced into asymmetric flagellar beating patterns enabling directional bias and controlled reorientation in response to environmental light gradients. This sensory architecture, in conjunction with quantitative encoding of light intensity, enables precise alignment of the cell’s swimming trajectory relative to the incident light. The eyespot that we were talking about typically comprises a localized aggregation of carotenoid-rich pigment granules and in certain cases, it functions as an optical diffraction grating [168]. However, beyond simple phototactic steering, microorganisms possess a spectrum of sophisticated photoresponses that couple signal transduction with flagellar dynamics. Moreover, they respond to environmental changes through a combination of phenotypic plasticity, physiological acclimation, and genetic adaptations. Short term environmental changes are often managed first by plastic responses such as altering their physiology and morphology [41,47,56,169]. Contrarily, over longer timescales, populations may undergo genetic adaptations or selection among strains with distinct ecotypes with dedicated functional traits [170–173]. Relevant phenotypic and physiological properties include cell size/morphology, lipid droplets/pigment contents, nutrient affinity, uptake kinetics, motility, thermal tolerance, and often colony formations. Consequently, comprehensive theoretical frameworks should be formulated accordingly to capture the relevant higher order photoresponses. They must account for intensity-dependent taxis, transient behavioral switching, and the coupling between sensory modulation and collective hydrodynamic instabilities to properly detect the cell distribution, plume formation, and large-scale flow structures. Where thermotaxis results in patterns emerging from the interplay of microorganism motility, thermal diffusion, and induced fluid flow [174,175], chemotaxis results in accumulation depending on the gradients of chemical concentrations and depending on the size they might have different level and mechanism of sensitivity towards the external chemical gradient [107,176,177]. Hence, we should always be cautious about emerging thermotactic and chemotactic effects in addition to gravitactic and phototactic responses while utilizing photosynthetic microbes in applications starting from heat transfer to even drug delivery. Different taxis mechanisms are discussed in more detail in the next section.

### *4.2. Taxis mechanisms*

Bioconvection resulting from microorganism collective movement goes back to the individual movement driven by different factors which necessitate considerable understanding of their swimming mechanisms. In this section, we discuss the organism taxis mechanism in response to different cues and we would first focus on gravity driven swimming which is gravitaxis and phytoplankton species have been comprehended to exhibit gravitaxis executing diel vertical migrations [161]. Swimming against (negative gravitaxis) or along the gravity (positive gravitaxis) may depend on the physiological state, time of the day or the season, cells' developmental phase, and response or adaptation to exogenous stressors. In the context of dynamics behind the swimming mechanisms, we would not discuss flagellar beating and other cell scale swimming mechanisms in this article as there are several articles that have reviewed this. On the contrary, we will briefly discuss the taxis mechanisms before delving deeper into the bioconvection and theoretical to experimental insights that establish the potential of bioconvection in future thermal management and flow manipulation strategies. The first taxis mechanism that we would discuss is directly influenced by gravitational field. In response to the gravity force, the sensing and response mechanism in microbes follow a series of steps and they are: perception, transduction, signal amplification, and response. The receptors that detect gravity signals often involve heavy crystals such as $BaSO_4$ and $SrSO_4$ initiating mechano-signal transduction chains to differentiate directions and local acceleration changes [148,178]. Moving forward, phototaxis refers to the directed movement of organisms toward or away from a light source to optimize light exposure for physiological benefits [159]. Planktons comprise diversified drifting organisms, including photosynthetic microorganisms and motile protists, many of which actively respond to light despite living in flow-dominated environments. Photo sensitive plankton possesses specialized light-sensing systems that enable directional detection of illumination more regarding which we discussed in sections **4.1** whereas this section is dedicated more towards the taxis mechanisms.

Besides gravitaxis and phototaxis, oxytactic bacteria swim up driven by the concentration gradient of oxygen while accumulating at the air-water surface. Oxytaxis can be considered as a specific type of chemotaxis where the cells may move in response to chemical gradients in general. In such situations, if the cells are negatively buoyant, an overturning instability emerges after they accumulate in a particular region ultimately resulting in bioconvective patterns. Other than the experimental observations, models have been formulated to study such behavior with linear stability analysis [44] and weakly nonlinear analysis [44]. The linear analysis by Hillesdon & Pedley [44] found that the instability is non-oscillatory in shallow chambers where the same is oscillatory in deep chambers with the critical wavenumber remaining non-zero. The main taxis types driving bioconvection are listed in **Box 3**:

***BOX 3:*** *Dominant taxis mechanisms are listed with a representative organism that shows the particular taxis mechanisms and their roles in bioconvection.*

| Taxis type | Mechanism | Key organisms |
|---|---|---|
| Gravitaxis [46,48,57,119,121,122] | Cells swim against gravity (negative gravitaxis) due to bottom-heaviness or statocysts | *Heterosigma akashiwo* |
| Gyrotaxis [45,46,48,59,105,108,109,179] | Bottom-heavy cells reorient toward downwelling via gravitational and viscous torque balance | *Heterosigma akashiwo*, *Chlamydomonas reinhardtii* |
| Phototaxis [48,99,168,180] | Directed motion toward (positive) or away from (negative) light via photoreceptors | *Euglena gracilis, Chlamydomonas reinhardtii* |
| Oxytaxis/Aerotaxis [44,46,48,95,181] | Motion up oxygen gradients via energy taxis or aerotactic sensors | *Bacillus subtilis*, *Escherichia coli* |
| Chemotaxis [46,48,138,140,176,182] | Response to chemical gradients (nutrients, toxins) | Bacteria (*Escherichia coli*), algae |

Nonetheless, microorganisms may often portray combinations of two or more taxis mechanisms while swimming in versatile environmental conditions. Starting from cell level swimming to collective level movement, countless attributes are there and to characterize them, we must look into the representative length scales. We dedicate our next section to discuss about crucial length scales in bioconvection.

### *4.3. Various scales in microbial dynamics and relevant non-dimensional numbers*

In the ocean, the inhomogeneity any turbulence carry may become insignificant to microbes, and they hardly feel the jerk while flowing with even the devastating waves. Interestingly, it will be naïve to consider that they are indifferent to laminar and turbulent behavior of flow [47,56]. Their migration behavior has been observed to vary significantly depending on the turbulence [56].

Where each microbe has size in micrometers, they astonishingly influence several meters in the ocean and lab scales [59,87]. The cell-cell interaction may take place in $10-100\ \mu\mathrm{m}$ range and we still don't know countless attributes of their communication strategies and how internal and external parameters may substantially influence their collective communication mechanisms. Several physiological and biomechanical factors are there that directly relate to the relative scales in the phenomena and here we aim to consider relevant scales that are substantial in dictating the complex dynamic attributes in bioconvection spanning more than six orders of magnitude.

**Cell scale (~10 µm):** Individual gyrotactic or phototactic swimming establishes taxis bias $\chi$ that leads to instabilities. *Chlamydomonas reorients within* $\sim 0.5-2\ \mathrm{s}$ to align with lights or gravity generating upward motion at velocity in the range of $100\ \mathrm{\mu m/s}$ [91]. Additionally, cell interacting with the fluid media may induce changes in rheological and hydrodynamic properties.

**Microscale (~100 µm):** Cell-cell hydrodynamic interactions via Jeffery orbits [183] and cell-boundary interactions [184] may induce polarity correlations decaying over $10-20$ times body lengths enhancing active diffusivity and stabilizing early instabilities within the system [100].
**Pattern scale (~1-10 mm):** Plumes and cells emerge with vertical velocities $100-500\ \mathrm{\mu m/s}$, vorticity concentrating at plume edges.

**Meso-scale (~1-30 cm):** Laboratory scale suspensions exhibit overturning circulation with pattern wavelength ($2-10\ \mathrm{mm}$), evolving into large-scale rolls that homogenize suspensions.
**Field scale (~1-10 m):** Natural scales sustain meter-scale convective plumes through persistent density-driven overturning at $\sim 10^6\ \mathrm{cells/mL}$, reorganizing biogeochemical landscapes.

**Kolmogorov scale and microbial activity:** Now, one important aspect is to understand characteristic bioconvection length scale relative to the Kolmogorov length scale which can constitute a critical threshold for interpreting microorganism-flow interactions underlying bioconvective phenomena. If the Kolmogorov length scale is much higher than the microorganism, then they would hardly experience the turbulence and if the swimming or collective forcing becomes comparable to the smallest eddies, it can disrupt or exploit the structure. In turbulent to transition regimes, energy carried at large scales is induced through inertial subranges prior to getting dissipated at the Kolmogorov scale. As the characteristic size and swimming induced disturbances of several microorganisms lie well within or below this dissipative range, their dynamics are mostly governed by local strain-rate and vorticity fields rather than inertial transport. In bioconvections, microscale interactions such as Jeffery-like rotational dynamics [183], gyrotactic torques, and viscous reorientation influence the onset of the instabilities by modulating hydrodynamic and diffusive attributes within the suspension. Therefore, the Kolmogorov scale provides a natural framework for mapping turbulence theory with active suspension mechanics enabling multiscale closure models that can connect energy dissipation rates to collective plume formations, pattern wavelength selection, and stability characteristics.

### *4.4. Flow coupling and rheological changes*

Flow coupling and rheological changes in bioconvection arise from a tight feedback loop between microorganism concentrations, motility, evolving density fields, and the surrounding fluid mechanics [48,108,185,186]. In bioconvection, as unstable density layers start to overturn, the sinking plumes must be balanced by lateral and returning flows that advect fluid (and cells) back into upwelling regions, closing a global overturning circulation in which cell transport, buoyancy, and vorticity are strongly coupled. At the suspension scale, this feedback modifies the effective rheology: the active stresses and self-generated shear my often lead to non-Newtonian behavior, apparent viscosity changes, and anisotropic normal stresses. From an engineering perspective, this means that classical constitutive relations must be augmented by active contributions tied to dimensionless groups such as the bioconvection Rayleigh number and swim Péclet number, which together control when this self-organized overturning emerges and how strongly it reshapes momentum and heat transport in the fluid. Therefore, we earlier took an attempt to carefully revisit relevant non-dimensional numbers that can become crucial to characterize and formulate such flow patterns with greater universality and perception. At this low Reynolds number inertial regime, the governing Stokes equations become kinematically reversible dictating that any reciprocal body deformation yields zero net displacement as formalized by Purcell's scallop theorem [187]. To achieve an effective self-propulsion, a micro-swimmer must execute a sequence of shape deformations that maps out a nonreciprocal, area-enclosing trajectory in its configuration phase space, thereby breaking time-reversal symmetry. The effective rheological properties depend on cell concentration, cell morphology, swimming mode, and the local shear environment [116,188,189]. The anomalous rheology stemming from the swimming characteristics lies in the force-dipole structure of the micro-swimmer flow field and the stresslet disturbances induced by the active particles [185]. In addition to the ability to self-propel, the translocation of swimming microorganisms involves another crucial factor which is the orientation decorrelation mechanisms while randomizing the orientations turning this into diffusive random walks. The effective swim diffusivity can be expressed as, $D_{eff} = {V_s}^2\tau_r/6$, where $\tau_r$ is the mean duration of a runlength, $l_r = V_s\tau_r$. In various swimmers, correlations have been found in pre- and post-tumble directions [190] and conditional probability has been introduced accordingly [191]. Moreover, other than swimmer transport properties, another critical factor is the disturbance flows induced by individual particles. The flow field generated by a swimmer depends on the body kinematics, surface stresses, and possible surface slip [188,192]. The disturbance flow induced can generally be expressed as,

$$\boldsymbol{u}(x) = \frac{1}{8\pi\mu}\int_{\partial V} \mathbf{J}(x - x_0)\cdot\boldsymbol{t}(x_0)\,dA_0 + \frac{1}{4\pi}\int_{\partial V} \boldsymbol{n}(x_0)\cdot\boldsymbol{K}(x - x_0)\cdot\mathbf{v}(x_0)\,dA_0 \qquad (9)$$

where $\partial V$ is the particle surface with outward normal $\boldsymbol{n}$, and $\boldsymbol{t} = \boldsymbol{n}\cdot\sigma$ is the surface traction expressed as a function of stress tensor $\sigma$ in fluid of viscosity $\mu$. The kernels $\mathbf{J}$ and $\boldsymbol{K}$ are given by the Oseen tensor and its symmetric gradient [193]. Restricting ourselves from delving deeper into the complexities of flow disturbance related intricate physical interactions, we would like to wrap this part up with mentioning about flow singularities where stokeslet represents the net force acting and stoke doublet represents the moment (symmetric part) and torque (antisymmetric part) applied on the fluids [185,193]. In parallel to theoretical modelling endeavors, [194,195] several experimental works are there where researchers have measured the disturbance flow induced by living microorganisms such as *C. reinhardtii* [196,197]. Additionally, nonzero Stokeslet may be present for nonneutrally buoyant swimming microorganisms while swimming against gravity [105,196].

The localized disturbance flows may directly counteract or enhance the forces of an external flow. For a non-Brownian slender particle in shear flow, it follows the orientation dictated by the Jeffery orbit with additional stochastic attributes involving forces like rotational diffusion and run-and-tumble dynamics disrupting the perfect balance. Moving forward, the active swimmers completely change the scene while introducing active stresslet component in addition to the passive components [185]. The pushers which pull themselves by pushing fluid behind them such as *E. coli* [198] or *Bacillus subtilis* [199] tend to align in the extensional quadrant of the applied shear, reinforcing rather than resisting the flow, thereby reducing the effective shear viscosity [198,199]. The effective viscosity of the suspension was experimentally observed to decrease by multiple times compared to the fluid without the micro-swimmers [199]. On the contrary, pullers which pull fluid from the font such as *Chlamydomonas reinhardtii* resist flow-induced rotation and thus increase the effective viscosity [200,201]. This pusher-puller asymmetry translates directly into thermal transport consequences as the rheological influences may often become crucial for sustaining the plume structures and consequently the heat redistributions. The effective viscosity of a bioconvective suspension involves non-linear feedback into the governing equations making the system qualitatively different from conventional thermal convection configurations. Recent numerical investigations have shown that the velocity fields in bioconvective suspensions are reduced by an increasing bioconvective Rayleigh number ($Ra_b$) and buoyancy ratio parameter reflecting the fact that as the microorganism concentration increases the convective velocities may get modulated in a non-linear, spatially heterogenous manner.

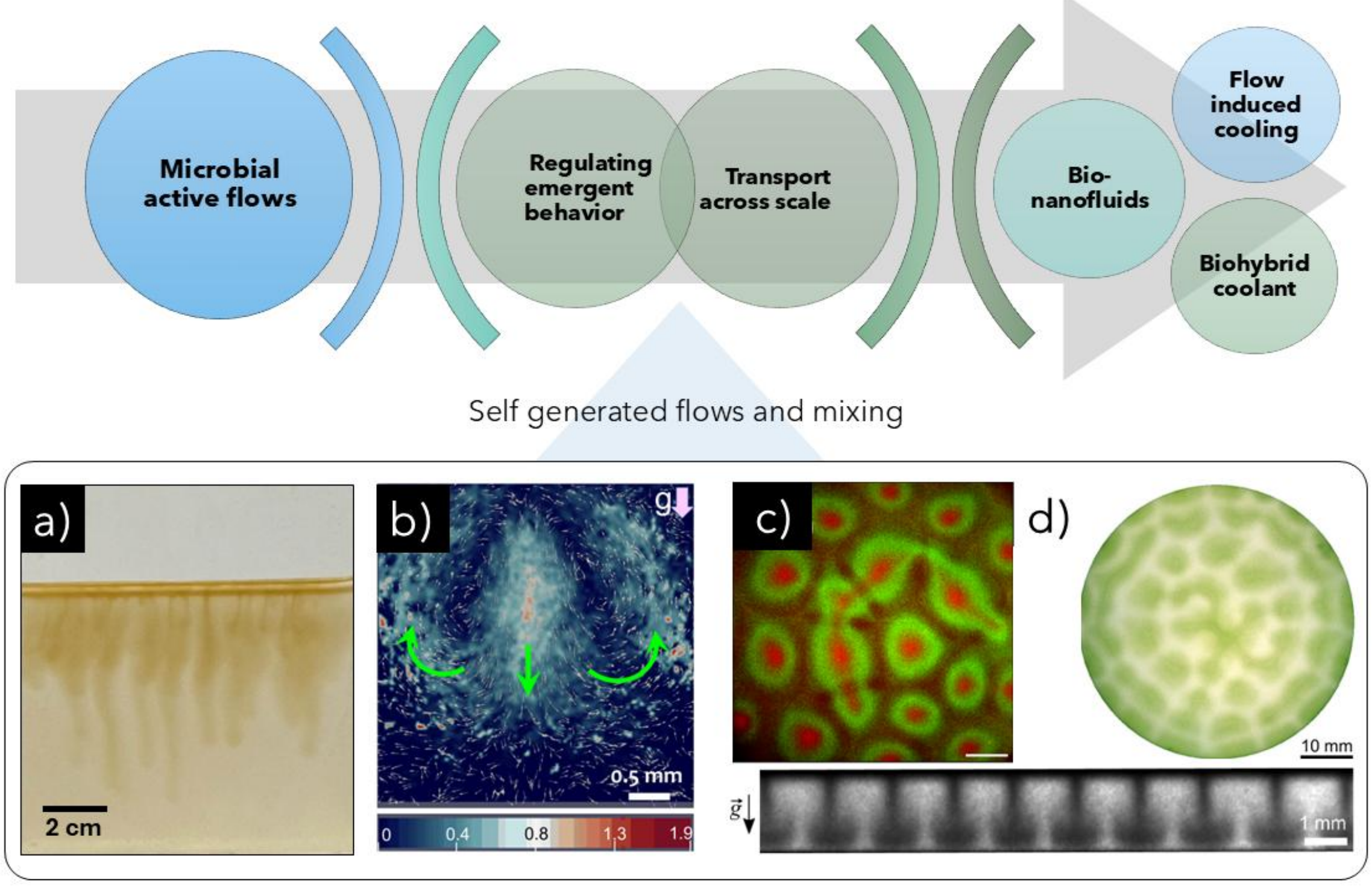

**Fig. 6. Features of microbial flows relevant to the development of bio-nanofluids and biohybrid coolants for future thermal management.** (a) Multiple bioconvective plumes generated by a gravitactic species. (b) Particle image velocimetry (PIV) of a stable plume in millifluidic confinement [41] (https://doi.org/10.48550/arXiv.2301.09550). (c) Dynamically segregated bioconvective patterns due to different phenotypic traits of 2 different species (scale bar $\sim 2$ mm) (Credit - Gallardo-Navarro *et al.*, 2025) [203] (https://doi.org/10.1038/s41467-025-56244-8). (d) Emergent patterns and coherent dynamics in living microorganisms controlled by light (Credit - Fragkopoulos *et al.*, 2025) [204] (https://doi.org/10.1073/pnas.2413340122).

Rheological responses of these active matter [185] often leads to non-trivial macroscale traces unlike any passive systems. Several non-intuitive transport properties such as accumulation near boundaries, enhanced diffusion, and rectification have been observed in the active matter propulsion. At these viscous domain, the active micro-swimmers often exploit drag using their propulsion mechanism [189]. Another astonishing observation was the superfluid like behavior of the bacteria suspension by Lopez *et al.* [202] where they observed the viscous resistance to shear to vanish, thus, microscopically, the activity of the swimmers is able to fully overcome the dissipative effects due to viscous loss**.** From here, we would take an attempt towards the experimental observations focusing on the collective behavior of active matter specifically organized patterns such as bioconvection.

## 5. Experimental observations in collective behavior

### *5.1. Species and mechanisms driving collective flows*

Bioconvection has been demonstrated experimentally across several major groups of motile microorganisms, and organizing the literature by species helps emphasize the broad biological diversity underlying this phenomenon. Among phytoplanktonic microalgae, some of the most widely used model systems are *Euglena gracilis* [205], *Chlamydomonas nivalis* [95], and *Chlamydomonas reinhardtii* [206]. Bioconvection is also well established in bacterial systems, particularly in *Bacillus subtilis* [99], where classic quantitative studies examined the onset conditions for bioconvective instability. In *Escherichia coli*, it has been observed that suspension depth, nutrient availability, and motility strongly influence the emergence and morphology of the resulting patterns. A third major group comprises ciliates and protozoa, including *Tetrahymena pyriformis* [207], *Tetrahymena thermophila* [207], and *Paramecium tetraurelia* [208], all of which have been used in experiments on dense suspensions, gravity-sensitive patterning, and the formation of propagating fronts.

The key mechanistic reason behind these large-scale dynamics is the taxis behaviour at a single cell level, that allows them to migrate away from gravity. In *C. reinhardtii*, for example, active migration driven by phototaxis can generate fluid motion sufficiently strong to stir the surrounding medium, and these flows have even been exploited as hydrodynamic tweezers for manipulating small floating objects [209]. In *Euglena gracilis*, another phototactic microorganism, strong illumination induces localized convection patterns, and the stability of these patterns has been shown to depend sensitively on the initial cell-density distribution [210]. For *Tetrahymena thermophila*, experiments conducted during parabolic flight under partial gravity conditions (< 1 g) revealed short-lived and shortened downward plumes together with reduced gravikinesis (thrust regulation), as resolved through dual-scale macro- and micro-imaging; notably, plume persistence decreased as gravitational acceleration was reduced [211]. In the marine alga *Effrenium voratum*, dense cultures exceeding normal cell concentrations exhibit gyrotaxis-driven aggregation near the air–water interface, followed by the formation of sinking plumes that enhance vertical nutrient mixing. Particle image velocimetry (PIV) measurements further showed that cell accumulation descend sequentially, and if it is observed in a confinement, roll-like structural patterns appear [212]. For the gravitactic alga *Heterosigma akashiwo*, Bearon and Grünbaum (2006) investigated bioconvection in a salinity-stratified system and found that, in weakly stratified fluids, plume structures as large as 20 cm could develop. They also showed that introducing a low-salinity surface layer promoted dense surface aggregation, suggesting a possible industrial strategy for concentrating cells [49]. A recent study showed that a mixed population of two microbial species can undergo dynamic segregation through bioconvection, driven by species-specific differences in motility. This segregation emerges due to hydrodynamic interactions rather than biochemical repulsion between the species [203]. These findings open the possibility of guiding material transport, segregation, and trapping in microbial consortia by exploiting differences in phenotypic traits.

Comparable mechanisms have also been documented in bacterial systems, where directional responses to environmental cues generate density inversions and collective motion. *Bacillus subtilis*, for instance, uses positive aerotaxis to accumulate at oxygen-rich air–liquid interfaces, thereby creating density instabilities that lead to plume formation and confer growth advantages in static cultures; interestingly, mutants lacking aerotaxis (Δ*aer*) still exhibit bioconvection through secondary cues [213]. In some bacterial suspensions, responses to quorum-sensing molecules or nutrient gradients can additionally promote collective plume formation, and in combination with aerotaxis these responses help sustain lattice-like patterns even in the absence of global gradients [214]. Manzano *et al.* (2020) reported that magnetotactic bacteria (MTB) accumulate near channel surfaces and self-organize into dynamic plumes that drive large-scale circulation. In that system, the applied magnetic field strength controlled plume dynamics, and the observations were successfully interpreted using hydrodynamic singularity models that balance magnetic alignment torques with cell–cell and cell–fluid interactions [214]. Hirota *et al.* (2017) similarly demonstrated chemotaxis-induced bioconvection in *Salmonella* using a serine capillary assay, where PIV measurements revealed convective flows arising from density instabilities near the chemoattractant source. Cell accumulation reached saturation before plume formation accelerated, confirming that the onset of convection was gravitationally triggered through local stratification [182].

### *5.2. Bioconvection in natural settings*

Bioconvection is not merely a laboratory-scale instability; it can also play a significant role in naturally stratified aquatic environments. One of the clearest natural examples comes from Lake Cadagno, a meromictic alpine lake, where Sommer *et al.* (2017) reported what they described as the first observation of bioconvection in a natural water body. Their study showed that the motile purple sulphur bacterium *Chromatium okenii* can generate a mixed layer approximately 0.3–1.2 m thick near the chemocline [87]. More recently, Di Nezio *et al.* (2023) linked this physical mixing directly to microbial ecology and metabolism. Based on field observations in the same lake, they argued that *C. okenii*-driven bioconvection homogenizes the bacterial layer and thereby modifies the local physicochemical environment as well as the eco-physiological performance of the phototrophic sulfur bacteria inhabiting it [215, 18]. Together, these studies show that bioconvection can operate at ecologically meaningful scales and may influence both thermofluidic transport as well as the microbial ecosystem structure in stratified natural waters [90].

### *5.3. Possibilities toward heat transfer and experimental challenges*

The ability of bioconvection to produce large-scale homogenization suggests that it is not simply a biological pattern-forming instability, but also a potentially important mechanism of thermofluidic transport. Bioconvection governs multiscale active transport by redistributing material within the local environment, from micron-sized cargo to molecular scale [90]. The resulting hydrodynamic transport can span length scales comparable to the size of the confinement. Because this transport is regulated by cellular phenotypic traits, it provides a potential route for controlling self-generated flows for heat-transfer applications. In addition, molecular transport and trapping depend on both molecular diffusivity and the location of release [90], suggesting that the position and intensity of a thermal source could be engineered to achieve regulated thermofluidic transport. Because bioconvective recirculation can redistribute heat, dissolved species, and suspended matter, it has attracted increasing interest in engineering contexts as a possible route to enhanced transport even involving porous medium [216]. Motivated by this idea, a growing body of theoretical and numerical work has examined bioconvective heat and mass transfer in idealized systems containing motile microorganisms, particularly in nanofluid and porous-medium formulations [175,181,217,218]. Recent examples include Hussain *et al.* (2024), who studied hybrid bioconvective flow with microorganism migration using the Buongiorno nanofluid framework under convective boundary conditions, and Rashad (2019), who analysed gyrotactic bioconvection around a heated cylinder [219]. Related studies have similarly reported that incorporating motile microorganisms can alter boundary-layer structure, stabilize suspensions, and, in certain parameter regimes, enhance heat transfer or mixing in porous cavities and other thermally driven flow configurations [179,181]. In Fig. 6, we demonstrate investigations focusing on bioconvection like collective flows, together with the relevant experimental and theoretical works. Both rheological interactions, and the collective flows including the bioconvective patterns in lab and natural settings can have direct implications towards future thermal management strategies.

Although the thermo-bioconvection studies remain limited to the numerical observations and are still far removed from immediate device-level implementation, they collectively support an emerging concept: if microbial activity can drive sustained recirculating flows across scales ranging from microscopic suspensions to meter-scale natural layers, then microorganism-induced convection may offer new opportunities for passive or low-energy mixing, thermal regulation, and transport enhancement in bioinspired fluidic systems. At the same time, significant experimental barriers remain before such ideas can be translated into practical heat-transfer technologies. First, bioconvective flows are highly sensitive to the physiological state of the cells. Parameters such as motility, concentration, nutrient availability, oxygen level, and light exposure all strongly influence whether a stable bioconvective pattern develops at all, making reproducibility much more challenging than in conventional forced or buoyancy-driven convection. Second, the flow is strongly affected by the external operating environment. Experiments on *Chlamydomonas* in horizontal tubes have shown that imposed through-flow can distort, fragment, or partially suppress bioconvective plumes, highlighting a key challenge for any device that requires both circulation and efficient heat exchange simultaneously [220]. Although recent photo-bioconvection studies demonstrate that light can provide precise, rapid, and reversible control over plume formation and global recirculation, implementing such control in a functional heat-transfer device would require careful optimization to balance optical forcing with cellular viability and long-term operational stability. Additionally, the dynamics of micro-swimmers within confined environment under the influence of physicochemical properties, external fields, and nanoparticles are yet to be revealed for future hybrid coolant or thermal management technologies. Under confinement, functionalisation of the surface may largely impact flow patterns and swimming behaviour which can serve as additional control parameters. Microorganisms can undoubtedly generate collective organised flow structures; however, demonstrating that such flow can be maintained, controlled, and quantified with sufficient reliability to deliver useful and scalable heat-transfer enhancement remains one of the major challenges in developing future sustainable biobased heat-transfer device. Fig. 7 shows a step-by-step observation and basic methodologies taken up where the motile microbes is performing bioconvection (Fig. 7.b-c) and how this can lead to redistribution of heat in a chamber (Fig. 7.d). Detailed velocity and flow fields can be extracted by imaging the phenomena and thermal imaging can be used to quantify the temperature fields homogenizing capability.

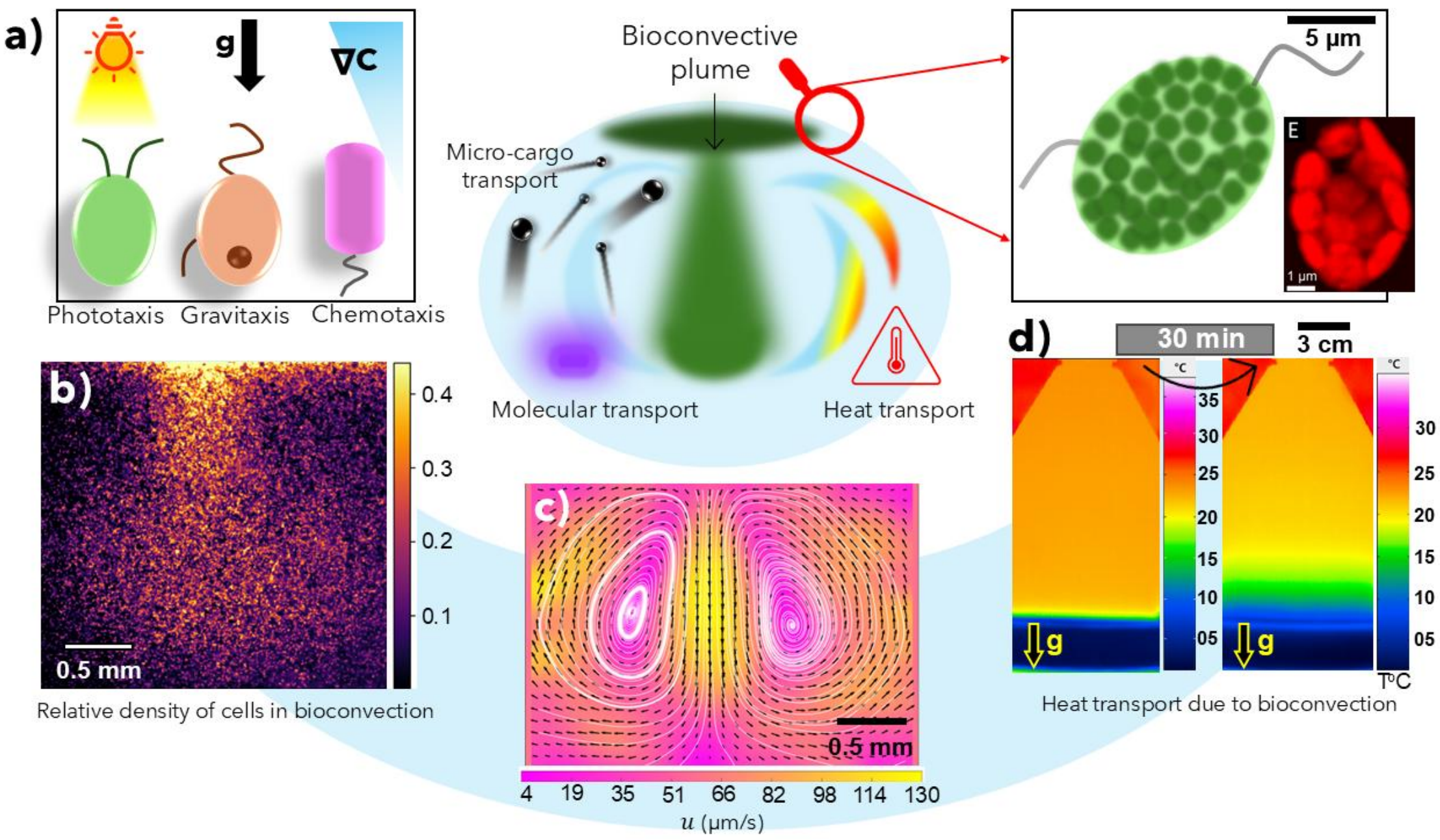

**Fig. 7. *Thermofluidics-on-chip* approach promises to harness bioconvective transport for thermofluidic applications.** (a) Different taxis mechanisms lead to the onset of bioconvection. (b)The bioconvective plumes observed at the laboratory [90] in the *thermofluidics on-chip* device is further investigated using image analysis and (c) The velocity field and streamlines of the bioconvective flow [90]. (d) Bioconvection mediated heat transfer, captured using a thermal camera. Faster homogenisation and mixing in temperature fields directly shows how bioconvection drives thermal transport. The inset on the top right shows single cell morphology and autofluorescence image of *H. akashiwo* [41].

## 6. Challenges and future directions in technology transfer

Harnessing microbial activity-mediated transport into practical applications in cooling or heat transport technologies will necessitate precise control, deeper understanding, and design strategies all of which are in an early stage presently. Depending on whether the "active" agent is a living organism (such as phytoplankton [47]) or a synthetic active material (such as Janus colloids or active gels/particles) [13,221–223] the level complexities may slightly vary from one case to another.

### *6.1. Challenges in harnessing active matter in real-life applications*

In industrial applications, the primary challenges associated with active matter can be broadly categorized into three areas: control and predictability, scalability and geometric constraints, and integration with existing infrastructure. As inherently nonequilibrium systems, active matter exhibits nonlinear and often chaotic dynamics, making it difficult to achieve reliable, stable, and predictable performance under practical operating conditions. A second major challenge concerns scalability. While many active matter phenomena have been successfully demonstrated at laboratory scales, it remains unclear whether their underlying mechanisms and performance will scale proportionally to larger, industrially relevant geometries. Addressing this gap requires a deeper understanding of scaling laws and the influence of complex geometries on collective dynamics. Finally, the successful deployment of active matter technologies depends on their seamless integration with existing engineering systems without compromising operational efficiency or reliability. Nevertheless, recent advances in microfabrication, sensing, materials science, computational modeling, and real-time control provide unprecedented opportunities to overcome these challenges. Future research should therefore focus on developing scalable, controllable, and application-specific active matter systems capable of translating laboratory-scale demonstrations into robust industrial technologies

- ***Cell physiology:*** Cells divide, age and die, spanning their growth curve. Beyond these, the cell suspension composition drifts over time, changing motility, tactic response, and transport properties [44]. Hence, maintaining stable performance requires continuous monitoring and proper management of growth phase, density, and cell fitness.

- ***Nutrients and optimal habitats:*** Organisms need nutrients, light, or chemical energy to survive and flourish. Providing this in industrial loop may offer challenges which needs to be addressed for innovating next generation of thermal systems.

- ***Biofouling:*** Living cells frequently attach to surfaces forming biofilms and trapping particulates. This can pose challenges, as biofilms could increase hydraulic resistance and reduce the heat-transfer coefficients over time. Maintaining cleanliness from time to time is therefore required.

On the other hand, for the synthetic active matter, the challenges may slightly vary, and they are summarized as below.

- ***Fuel and actuation:*** Most of the synthetic swimmers need some chemical fuels (such as hydrogen or bubbles [224,225]), light at specific wavelengths [226], or electric/magnetic fields [227,228]. Continuous supply of the resources with uniformity and safety can be sometimes energy-intensive and may negate the efficiency gains.

- ***Materials compatibility:*** Catalytic and metallic particles may often corrode the system components, poison catalyst downstream, or even degrades seals and coatings [228,229].

- ***Degradation and stability:*** Synthetically active particles may often sediment, agglomerate, or even lose activity over time depending on the precise fabrication or environmental uncertainty.

- ***Cost and recyclability:*** Engineered particles mostly have size at the micrometer and nanometer ranges which requires highly sophisticated nanoengineering equipment and fabrication methods making the whole process costlier than the traditional coolant. Other issues are their recyclability, sustainability, and careful disposal as they often contain toxic metals or heavy metals.

*6.2 Adaptability*

Quantifying the adaptability of both living and synthetic active matter for industrial applications remains a critical milestone toward their practical deployment. Achieving this requires a careful assessment of their capabilities while identifying and mitigating the associated limitations and risks. In this section, we briefly examine the key advantages, challenges, and potential risks of these active matter systems, highlighting the essential considerations that must be addressed before their successful integration into industrial and commercial technologies.

*6.2.1. Living matter*

**Advantage:** Microbes can adapt to changing conditions, evolve improved traits, and self-repair populations after perturbation. This is a unique strength for long-lived systems exposed to variable loads or climates.

**Risk:** The same adaptability can drift performance away from design values (e.g. evolution toward less motile, biofilm-forming phenotypes that are energetically cheaper for the cells). Without evolutionary and ecological control strategies, an initially well-behaved active coolant can become less effective or more problematic over time.

*6.2.2. Synthetic matter*

**Advantage:** Once designed, synthetic active particles are typically chemically and behaviorally stable; they do not mutate or reproduce. This can make performance more predictable over time compared with living systems.

**Limitation:** They lack intrinsic adaptation. If the operating conditions change (temperature range, geometry, load), particles cannot “learn” or evolve; any new regime may require redesign or a different formulation.

Ultimately, combining these two components in a hybrid element may often reduce the individual limitations and can offer higher efficiency when employed in applications. Moreover, the knowledge gathered while exploring the biocomponents and their interactions may inspire novel synthetic component designs which makes this discussion more fruitful.

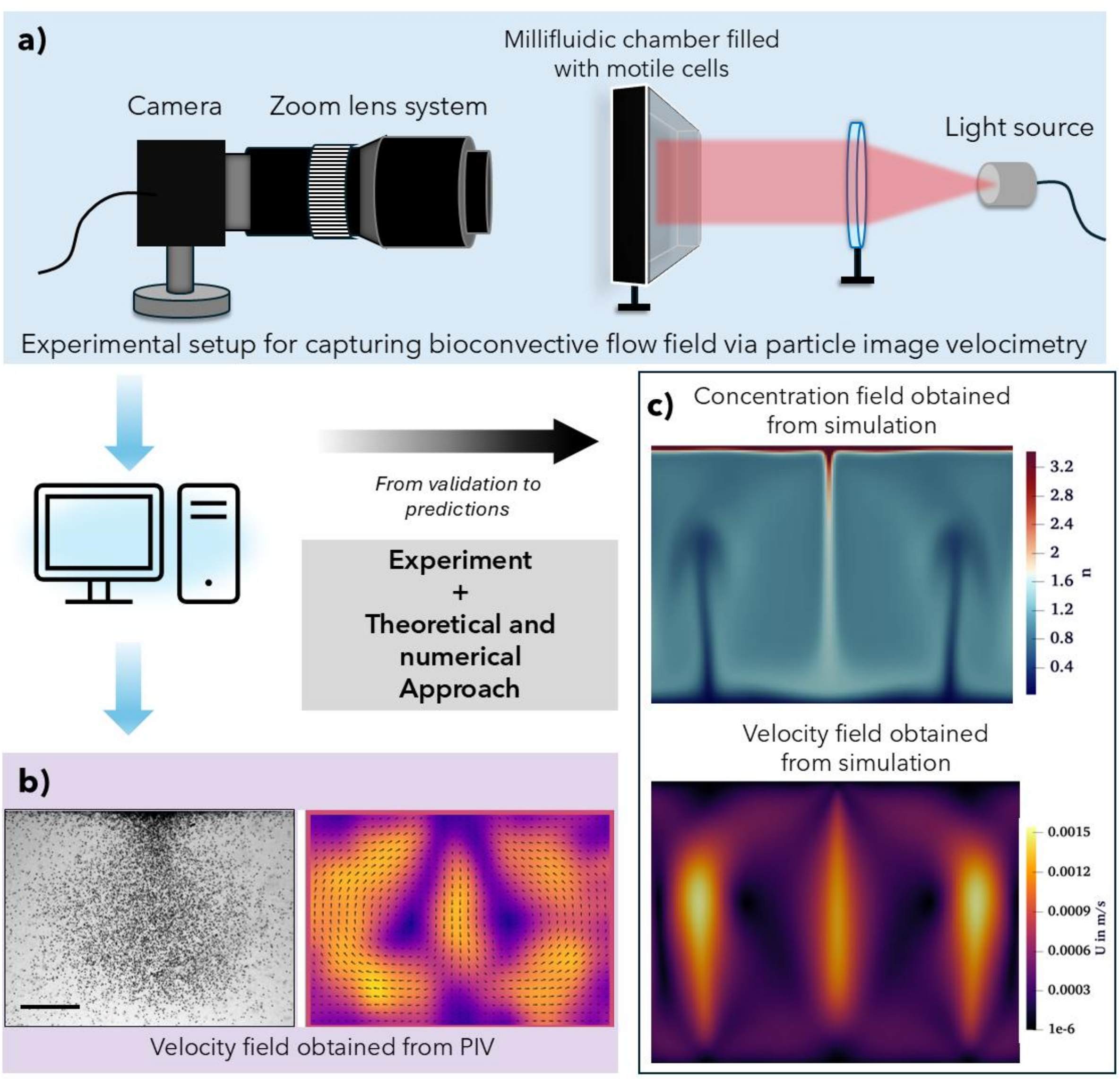

**Fig. 8. Toward *Living Thermofluidics*.** Integrating numerical framework with the experimental observation currently ongoing at the Physics of Living Matter Group, University of Luxembourg (Source: *Physics of Living Matter Group, University of Luxembourg*). (a) The experimental system to visualize collective scale behavior in a vertical setup. (b) Imaging bioconvective plume (scale bar $\sim 0.5$ mm) and the emerging flow field computed via PIV [90]. (c) Concentration and velocity fields obtained from the CFD integrated simulation for the bioconvection.

## 7. Conclusions

We envision a new generation of sustainable thermal management technologies that exploit living microorganisms and bio-hybrid systems either as standalone cooling solutions or as complementary components to conventional cooling methods. A key question, however, is whether microorganisms can remain viable under the elevated temperatures encountered in practical heat transfer systems. Rather than being directly exposed to high-temperature environments, microorganism-based heat transfer strategies can be implemented in secondary cooling loops, where operating temperatures typically remain within 20–40 °C, thereby enabling significant energy savings by reducing the need for forced cooling.

Furthermore, synergistic interactions between bioconvective microorganisms and engineered materials offer exciting opportunities for the development of next-generation synthetic and bio-based coolants. Recent advances in micro- and nanofabrication—including photolithography, soft and two-photon 3D lithography, reactive-ion etching, laser micromachining, and surface functionalization—now enable the fabrication of precisely engineered nanostructured surfaces, porous architectures, and tunable wettability that can dynamically interact with microorganisms, influencing both thermal boundary layers and hydrodynamic flow patterns. Likewise, optical and optogenetic techniques provide powerful tools for the precise manipulation of phototactic microorganisms, allowing light-controlled regulation of collective flow structures.

A fundamental prerequisite for these technologies is the quantitative characterization of thermofluidic performance and the influence of thermal cues on microbial behavior, particularly at the collective level where bioconvective plumes govern heat transfer and hydrodynamic transport. Integrated micro-/nanoheaters, high-resolution thermometry, and embedded sensing platforms can facilitate closed-loop temperature control while simultaneously correlating microbial distributions with local heat flux. Toward this goal, the *Physics of Living Matter Group* at the University of Luxembourg has recently developed integrated experimental and numerical frameworks to investigate these mechanisms, as illustrated in Fig. 8(a–c).

In conclusion, this comprehensive review demonstrates that bioconvective hydrodynamics encompasses a rich spectrum of multiscale phenomena, at the interface of active biological matter, nonlinear fluid dynamics, and self-organized pattern formation. This interdisciplinary nexus spans fundamental biology, soft matter physics, and thermal engineering. The complex interplay between organismal taxis, cell-scale hydrodynamic interactions, and emergent mesoscale instabilities gives rise to a wide range of flow structures, from Kolmogorov-scale active turbulence to system-scale convective overturning, many aspects of which remain incompletely understood despite decades of investigation. These phenomena nevertheless hold significant promise for next-generation thermal technologies. In the context of rising global energy demands and sustainability constraints, bioconvection offers a form of self-pumped convective enhancement ($Nu = 3 - 10$ at negligible metabolic cost), positioning it as a compelling candidate for addressing thermal bottlenecks in energy-intensive systems, including data center cooling and industrial heat management. This synthesis bridges experimental observations with quantitative scaling analyses, elucidating potential adaptation pathways while critically assessing challenges related to controllability, stability, and scalability of active matter systems. By consolidating these insights, we outline a pathway toward the engineering of bioconvection-based thermal management strategies, transforming biologically driven pattern formation into practical, scalable, and sustainable heat transfer technologies.

## Glossary

| Symbols | Definitions | Symbols | Definitions |
|---|---|---|---|
| $n$ | concentration of the cells | μ | viscosity |
| g | gravitational acceleration | $L_{ch}$ | characteristic length |
| ρ | density | $T$ | temperature |
| $V_{cell}$ | cell volume | β | thermal explansion coefficient |
| $D$ | diffusion coefficient | ω | Angular velocity of the cell |
| $V_S$ | swimming velocity | $k$ | thermal conductivity |
| $B$ | reorientation time | $Nu$ | Nusselt number |
| $Pr$ | Prandtl number | $Pe$ | Peclet number |
| $Ra_b$ | Bioconvective Rayleigh number | $Ra_T$ | Thermal Rayleigh number |
| $Pe_{swim}$ | Swimming Peclet number | $\boldsymbol{u}$ | Velocity vector |
| $c_P$ | Specific heat capacity | α | Thermal diffusivity |
| $h$ | convective heat transfer coefficient | $k_{eff}$ | Effective heat conductivity |
| $C_H$ | Center of the hydrodynamic Stress | $C_B$ | Center buoyancy |
| $F_P$ | Propulsive force due to flagella beating acting along the long axis | $F_D$ | Drag force through the center of Hydrodynamic stress |